\documentclass[prc,aastex,amsmath,amssymb,twocolumn,floatfix,aps,showpacs]{revtex4-2} 

\usepackage{graphicx} % Include figure files
\usepackage{amsmath}
\usepackage{amssymb}
\usepackage{textcomp}
\usepackage{xcolor}
\definecolor{myblue}{rgb}{0.0, 0.0, 0.6}
\usepackage{hyperref}
\hypersetup{
  colorlinks = true,
  citecolor  = myblue,
  linkcolor  = myblue,
  urlcolor   = myblue
}
\usepackage{multirow}
\begin{document}
\title{
Modeling a Framework for Comprehensive Analysis of $\boldsymbol{p^\uparrow\!\!\!{A}}$ Transverse Analyzing Power Measurements at the RHIC HJET
}
\author{A.~A.~Poblaguev}
\email{poblaguev@bnl.gov}
\affiliation{
  Brookhaven National Laboratory, Upton, New York 11973, USA
}

\date{November 2, 2026}

\begin{abstract}
  We present a model for calculating the forward elastic analyzing power $A_\text{N}^{pA}(t)$ in proton-nucleus scattering, based on the Glauber approximation and incorporating soft-photon exchange corrections. The model is tested against 100 GeV HJET measurements for several nuclei, including Au, Ru, Zr, Al, O, C, and the deuteron. The calculations use Woods-Saxon parameters to describe the nuclear density distributions, as well as the proton-proton and proton-neutron hadronic spin-flip amplitudes, with the latter constrained by a Regge fit to HJET $pp$ data. Using the nominal input parameters, the calculations provide a reasonable description of the measured analyzing powers, although further refinement is needed to achieve a satisfactory fit.  Allowing the input parameters to vary in the fits substantially improves the agreement with the data, yielding effective parametrizations of $A_\text{N}^{pA}(t)$ in terms of three or four parameters, depending on the nucleus. Notably, the nominal values of the imaginary part of the effective proton-nucleus spin-flip parameter $r_5^{pA}$ agree well with the fitted values for all nuclei considered.
  Subject to the limitations of the present model, this observation concerning $\operatorname{Im}r_5^{pA}$ is consistent with the proton-proton spin-flip amplitudes measured at HJET and supports the interpretation of a non-zero Pomeron spin-flip coupling.
  The results presented in this paper also motivate further investigation of the energy dependence of the proton-nucleon spin-flip amplitude using HJET measurements over a broad range of beam energies, with potential applications to proton-carbon and \(^3\)He polarimetry at the Electron-Ion Collider.

\end{abstract}

\maketitle

 \section{Introduction}

The Polarized Atomic Hydrogen Gas Jet Target (HJET)\,\cite {Zelenski:2005mz, Poblaguev:2020qbw} was designed to provide an absolute measurement of the transverse polarization of the proton beam in the Relativistic Heavy-Ion Collider (RHIC) Spin Program \cite{Bunce:2000uv}. In HJET measurements, a high-energy, vertically polarized proton beam scatters from a vertically polarized proton jet target, and the recoil protons are detected in left-right symmetric Si detectors. By measuring the recoil proton asymmetries relative to the beam and jet spin directions, $a_\text{beam}$ and $a_\text{jet}$, respectively, the beam polarization can be determined using the well-known jet polarization, $P_\text{jet}\approx96\pm0.1\%$,
\begin{equation}
  P_\text{beam} = \frac{a_\text{beam}}{a_\text{jet}}\times P_\text{jet}.
\end{equation}

Although the original HJET design requirement was to determine the proton beam polarization with an accuracy of 5\% \cite{Bunce:2000uv}, the upgrade of the recoil spectrometer and Data Acquisition System in 2015 \cite{Poblaguev:2015ptq} reduced the systematic uncertainties in the polarization measurements \cite{Poblaguev:2020qbw} to
\begin{equation}
  \sigma_P^\text{syst} / P_\text{beam} \approx 0.5\%,
\end{equation}
enabling precise measurements of the analyzing power as a function of the squared momentum transfer $t$,
\begin{equation}
  A_\text{N}(t) = a_\text{jet}(t) / P_\text{jet}.
\end{equation}

The elastic single-spin $A_\text{N}^{pp}$ and double-spin $A_\text{NN}^{pp}$ proton-proton analyzing powers have been measured \cite{Poblaguev:2019saw} in the Coulomb-Nuclear Interference (CNI) region,
\begin{equation}
  0.0013 < -t  < 0.018\,\text{GeV}^2,
  \label{eq:CNI}
\end{equation}
at two beam energies, 100 and 255 GeV.

For both observables, nonzero hadronic spin-flip contributions were reliably isolated. A Regge fit to the results indicates nonzero spin-flip Pomeron couplings for both amplitudes. Notably, this result disagrees with recent theoretical evaluations \cite{Ewerz:2016onn,*Hagiwara:2020mqb,*Sawasdipol:2023gro} of the spin-dependent Pomeron contribution.

\begin{figure*}[t]
  \begin{center}
    \hfill
    \includegraphics[width=0.90\columnwidth]{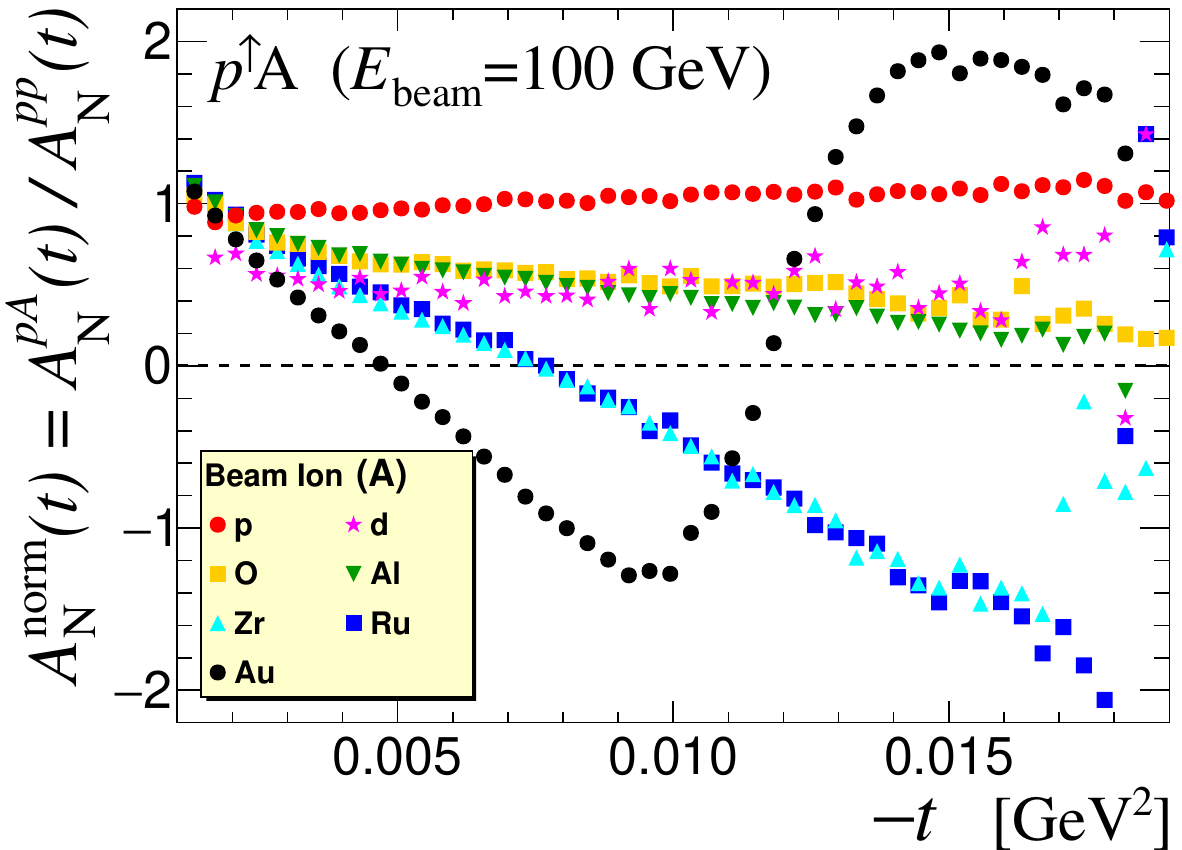}
    \hfill %\vspace{0.10\columnwidth}
    \includegraphics[width=0.90\columnwidth]{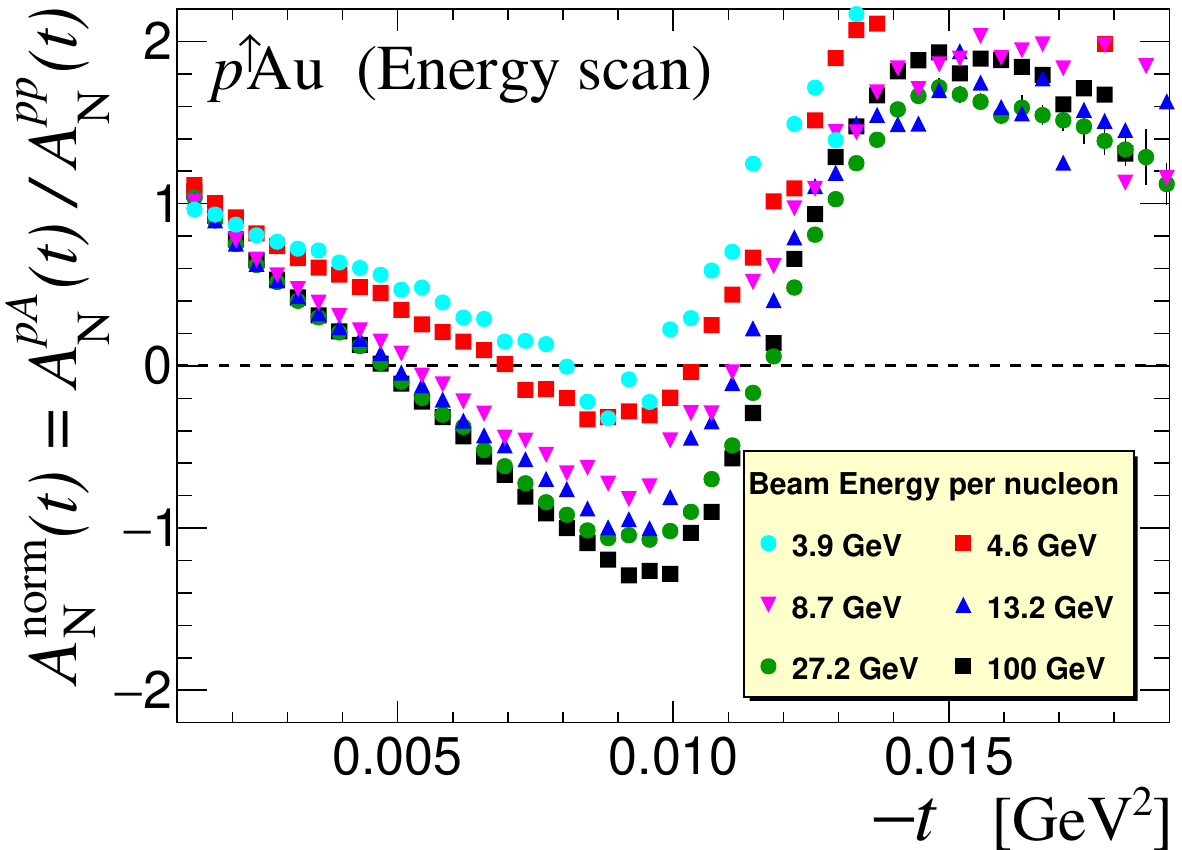}
    \hfill
  \end{center}
  \vspace{-1em}
  \caption{
    Preliminary results \cite{Poblaguev:2023stj} for dependence of the proton-nucleus elastic analyzing power $A_\text{N}^{pA}(t)$ on the beam ion (left) and beam energy (right). The measured analyzing powers are normalized to the proton-proton value calculated for $E_\text{beam} = 100\,\text{GeV}$, assuming no hadronic single-spin flip ($r_5 = 0$).
    \label{fig:pA}
  }
\end{figure*}

During the 2015 proton-gold and proton-aluminum runs at RHIC, the ion beams were passed through the HJET scattering chamber. HJET's good performance during these runs enabled the experimental determination of the proton-nucleus ($p^\uparrow A$) single-spin analyzing power, $A_\text{N}^{pA}(t)$, for several nuclei, as shown in Fig.\,\ref{fig:pA}. These measurements were routinely performed in parasitic mode (i.e., without disturbing RHIC experiments) during heavy-ion runs.

The primary goal of this paper is to develop a theoretical framework for calculating the transverse $p^{\uparrow}A$ analyzing power, thereby establishing a standardized parametrization of the $A_\text{N}^\text{pA}(t)$ measured at HJET. Such a parametrization is essential for completing the analysis of the HJET $A_\text{N}^{pA}$ data.

Although the HJET measurements involved nuclear beams scattering from a polarized proton target, Lorentz invariance allows the process to be interpreted, as it is in this paper, as the scattering of a polarized proton from a nuclear target.

\section{Forward elastic $\boldsymbol{p^{\uparrow}\!\!\!p}$ and $\boldsymbol{p^{\uparrow}\!\!\!\!A}$ analyzing power} 

In general, the elastic $p^\uparrow A$ transverse analyzing power is given by
\begin{equation}
  A_\text{N}(t) = \frac{2\,\operatorname{Im}\left(F_\text{sf}F_\text{nf}^*\right)}%
  {|F_\text{nf}|^2 + |F_\text{sf}|^2},
  \label{eq:AN}
\end{equation}
where the nonflip (nf) and spin-flip (sf) amplitudes for forward elastic $p^\uparrow A$ scattering can be written\,\cite{Vanzha:1972rps} in the Born approximation as sums of hadronic (N,S) and electromagnetic (C,M) contributions:
\begin{align}
  F_\text{nf}(\boldsymbol{q}) &= \frac{\sigma_\text{tot}}{4\pi} \left[%
    (i+\rho)\hat{f}_N(q^2)-q_c^2\hat{f}_C(q^2)
  \right],
  \label{eq:Fnf}
\\
  F_\text{sf}(\boldsymbol{q}) &= \frac{\sigma_\text{tot}}{4\pi} \left[%
    r_5\hat{f}_S(\boldsymbol{q}) - \frac{\kappa_pq_c^2}{2}\hat{f}_M(\boldsymbol{q})\right].
  \label{eq:Fsf}
\end{align}
Here, $\boldsymbol{q}$ is the transverse momentum of the scattered proton, with $q^2 \approx -t$, and $q_c^2 = -t_c = 8\pi\alpha Z/\sigma_\text{tot}$, where $Z$ is the nuclear charge. The parameters $\kappa_p = 1.793$, $\sigma_\text{tot}$, $\rho$, and $r_5$ represent, respectively, the proton anomalous magnetic moment, the total $pA$ cross section, the ratio of the real to imaginary parts of the hadronic amplitude, and the hadronic spin-flip parameter.

Assuming exponential form factors expressed in terms of $q^2$, the four basic amplitudes entering Eqs.\,\eqref{eq:Fnf} and \eqref{eq:Fsf} can be written as
\begin{align}
  f_N(q) = c_N\!\times\!\hat{f}_N(q)%
  &= \frac{(i+\rho)\sigma_\text{tot}}{4\pi} \times e^{-B_Nq^2/2},
  \\
  f_S(\boldsymbol{q}) = r_5c_S\!\times\!\hat{f}_S(\boldsymbol{q})%
  &= r_5\frac{\sigma_\text{tot}}{4\pi} \times%
  \frac{\boldsymbol{n\!\cdot\!q}}{m_p} e^{-B_Sq^2/2},
  \\
  f_C(q) = c_C\!\times\!\hat{f}_C(q)%
  &= -2\alpha Z \times \frac{e^{-B_Cq^2/2}}{q^2},
  \label{eq:fC} \\
  f_M(\boldsymbol{q}) = c_M\!\times\!\hat{f}_M(q)%
  &= -\kappa_p\alpha Z \times
  \frac{\boldsymbol{n\!\cdot\!q}}{m_p}\frac{e^{-B_Mq^2/2}}{q^2},
\end{align}
where $\boldsymbol{n}$ is a unit vector perpendicular to both the proton beam momentum and its spin, $m_p$ is the proton mass, and $Z=1$ for a proton target.

For 100\,GeV $p^{\uparrow}p$ scattering, the amplitude parameters are well known: $\sigma_\text{tot}^{pp}\!=\!38.39\,\text{mb}$~\cite{Fagundes:2017iwb}, $\rho^{pp}\!=\!-0.079$~\cite{Fagundes:2017iwb},
$t_c^{pp}\!=\!-1.85\times10^{-3}\,\text{GeV}^2$,
$B_S^{pp}\approx B_N^{pp}=11.2\,\text{GeV}^{-2}$\,
\cite{PhysRevLett.31.1088,*PhysRevLett.31.1367},
%\cite{Bartenev:1973jz,*Bartenev:1973kk}
$B_M^{pp}\approx B_C^{pp}=\frac{2}{3}\langle{r_p^2}\rangle=12.1\,\text{GeV}^{-2}$ \cite{Mohr:2025codata}, \hfill and\\
$r_5^{pp}\!=\!0.0164\!-\!i\,0.0053$\,\cite{Poblaguev:2019saw},

In the following calculations, we will use amplitudes in impact-parameter space, obtained by Fourier transformation:
\begin{equation}
  \gamma(\boldsymbol{b})=\frac{1}{2\pi}\int{%
    d^2{\boldsymbol{q}}e^{i\boldsymbol{q\cdot b}}f(\boldsymbol{q})
    \label{eq:Fourier}
  }
\end{equation}
For the exponential form factors, these amplitudes can be calculated analytically\,\cite{Poblaguev:2025vtd}:
\begin{flalign}
  \gamma_N(b)=c_N\hat{\gamma}_N(b),
  &\quad\hat{\gamma}_N(b)=\frac{e^{-b^2/2B_N}}{B_N}, 
  \label{eq:gN} \\
  \gamma_S(b)=c_S\hat{\gamma}_S(b),
  &\quad\hat{\gamma}_S(b)=i\frac{\boldsymbol{n\!\cdot\!b}}{m_p}\,%
    \frac{e^{-b^2/2B_S}}{B_S^2}, 
 \label{eq:gS} \\
  \gamma_M(b)=c_M\hat{\gamma}_M(b),
  &\quad\hat{\gamma}_M(b)=i\frac{\boldsymbol{n\!\cdot\!b}}{m_p}\,%
    \frac{1\!-\!e^{-b^2/2B_M}}{b^2}, 
   \label{eq:gM} \\
    \chi_C(b)=c_C\hat{\chi}_C(b),
    &\quad \nonumber
\end{flalign}
\vspace{-2\baselineskip}
\begin{flalign}
  & &\hat{\chi}_C(b)=\frac{1}{2}
  \left[\ln\left(\frac{b^2}{2B_C}\right) + E_1\left(\frac{b^2}{2B_C}\right)\right],  \label{eq:chiC}
\end{flalign}
where $E_1(z)$ is the exponential integral function. The Coulomb eikonal phase, $\hat{\chi}_C(b)$, is given in the massless-photon limit in Refs.\,\cite{Poblaguev:2025vtd,Poblaguev:2026wmb} (see Section\,\ref{sec:Coulomb} for more details).

For the $pp$ amplitudes, we will use the following notation:
\begin{align}
  &\gamma_N(b)\equiv\gamma_N(b,B_N),\quad\hat{\gamma}_N(b)\equiv\hat{\gamma}_N(b,B_N), \nonumber \\
  &\quad\hat{\gamma}_N(b,B)=\frac{e^{-b^2/2B}}{B},
\end{align}
with analogous definitions for $\gamma_S$, $\gamma_M$, and $\chi_C$.

For high-energy forward elastic proton-proton scattering, a compact expression for the transverse analyzing power $A_\text{N}(t)$ is well known\,\cite{Kopeliovich:1974ee,Buttimore:1978ry,Buttimore:1998rj}:
\begin{equation}
  A_\text{N}(t) = \frac{\sqrt{-t}}{m_p}\frac{\left(\kappa_p-2\operatorname{Im}r_5\right)t_c/t - 2\operatorname{Re}r_5}
  {(t_c/t)^2-2(\rho+\delta_C)t_c/t+1}.
  \label{eq:ANpp}
\end{equation}
Here, $\delta_C(t)$ denotes the Coulomb phase \cite{Cahn:1982nr, Kopeliovich:2000ez},
\begin{equation}
  \delta_C(t) \approx -\alpha Z
  \left[\gamma + \ln{\frac{(B_N+B_C)|t|}{2}}\right]\sim 0.02,
    \label{eq:deltaC}
\end{equation}
where $\gamma=0.5772\dots$ is the Euler-Mascheroni constant.

Although Eq.\,\eqref{eq:ANpp} omits some small corrections that are nevertheless non-negligible for precise experimental data analysis, the resulting analyzing power is in reasonably good agreement with measurements at RHIC energies.

Structurally, the analyzing power for $p^{\uparrow}A$ scattering has the same form as that for $p^{\uparrow}p$ scattering. However, for nuclei with large mass number $A$ and charge $Z$, the amplitudes exhibit a strong dependence on $q^2$, even within the CNI region \eqref{eq:CNI}. Furthermore, Coulomb corrections, which depend exponentially on $i\alpha Z$, are sufficiently large that they must be calculated to all orders in the fine-structure constant $\alpha$. In addition, the ultra-peripheral magnetic-photon correction to the hadronic nonflip amplitudes generates \cite{Kopeliovich:2021rdd,Kopeliovich:2023xtu} an effective spin-flip amplitude that is not included in Eq.\,\eqref{eq:ANpp} and may drastically alter the $q^2$ dependence of the analyzing power.

A framework for the theoretical interpretation of experimentally determined proton-nucleus analyzing powers was developed in Ref.\,\cite{Kopeliovich:2023xtu}. Here, we will largely follow this approach, with some modifications. In particular, the Coulomb correction will be evaluated to all orders in $\alpha Z$.

\section{Glauber approach for the $\boldsymbol{p^{\uparrow}\!\!\!\!A}$ amplitudes}

In the Glauber approach, the elastic scattering amplitude $f_N^{pA}(q)$ is expressed \cite{Glauber:1970wr} in terms of the nuclear profile function $\Gamma_A(b)$ as
\begin{equation}
  f_N^{pA}(q) = \frac{1}{2\pi}\int{
    d^2\boldsymbol{b} e^{-i\boldsymbol{q\cdot b}} i\Gamma_A(b)
  }.
\end{equation}
If we neglect correlations among the nucleon positions and write the many-body density function of the nuclear ground state in the factorized form $\prod \rho_p(r_i)$, then, assuming $A\gg1$, we obtain
\begin{equation}
  \Gamma_A(b) = 1-e^{i\chi_A(b)} = 1- \bigg[ 1+i\chi_{pN}(b)\bigg]^A
  \label{eq:GA}
\end{equation}

In Ref.\,\cite{Kopeliovich:2023xtu}, the proton-nucleon ($pN$) phase was evaluated as
\begin{equation}
  \chi_{pN}(b) = c_N\frac{2\pi T^h_A(b)}{A},
\end{equation}
where
\begin{equation}
 \frac{2\pi T_A^h}{A}(b) = \hat{\gamma}_N^{pN}(b) = \int{
   d^2\boldsymbol{s}\hat{\gamma}_N(s)
   \frac{T_A(\boldsymbol{b}-\boldsymbol{s})}{A},
   \label{eq:TAh}
 }
\end{equation}
is the convolution of the nuclear thickness function
\begin{equation}
  T_A(b) = A\int_{-\infty}^\infty{dz\,\rho(b,z)}
  \label{eq:TA}
\end{equation}
with the reduced $pp$ elastic amplitude $\hat{\gamma}_N(b)$. In Eq.\,\eqref{eq:TA}, the nuclear density distribution $\rho(r)$ is normalized to unity, i.e., $\int{d^3\boldsymbol{r}\,\rho(\boldsymbol{r})}=1$. Parametrization of $\rho(r)$ used in this paper, is described in Appendix\,\ref{sec:WS}.

For a nucleus with a Gaussian density distribution,
\begin{equation}
  \rho_A(r)=\left(\sqrt{2\pi}B_A\right)^{-3}e^{-r^2/2B_A^2},
\end{equation}
the thickness function is
\begin{equation}
  T_A(b) = \frac{A}{2\pi B_A}e^{-b^2/2B_A} = \frac{A}{2\pi}\hat{\gamma}_N(b,B_A),
\end{equation}
where $\hat{\gamma}_N(b,B_A)$ is the same function as that given in Eq.\,\eqref{eq:gN}, but with $B_N$ replaced by $B_A$.

In this case, interpreting Eq.\,\eqref{eq:TAh} as a convolution of the proton-proton $\gamma_N(b)$ amplitude with the nuclear thickness function, one obtains
\begin{equation}
  \hat{\gamma}_N^{pN}(b) = \hat{\gamma}_N(b,B_N+B_A).
  \label{eq:gNpN}
\end{equation}

This result implies that $\gamma_N^{pN}(b)= c_N\hat{\gamma}_N^{pN}(b)$ can also be considered an effective proton-nucleon amplitude for $pA$ scattering when the nuclear density $\rho(r)$ is not Gaussian. Consequently, approximating the proton-nucleus amplitude $\gamma_N^{pA}$ by multiple scattering of the beam proton off the nucleons in the nucleus (see Eq.\,\eqref{eq:GA}), we obtain
\begin{equation}
  \gamma_N^{pA}(b) = i\left\{1-\left[1+i\gamma_N^{pN}(b)\right]^A\right\},
  \label{eq:gNpA}
\end{equation}
and $f_N^{pA}(q)$ can be obtained by inverse Fourier transformation of $\gamma_N^{pA}(b)$.

If the amplitude $f(\boldsymbol{q})$ in Eq.\,\eqref{eq:Fourier} is given in the Born approximation, its Fourier transform should be interpreted as an eikonal phase. For example, the Born approximation for the Coulomb amplitude \eqref{eq:fC} is known from quantum electrodynamics. Consequently, $\chi_C(b)$ in Eq.\,\eqref{eq:chiC} is the eikonal phase, whereas
\begin{equation}
  \gamma_C(b)=i\left[1-e^{i\chi_C(b)}\right]
\end{equation}
is the corresponding amplitude in the impact-parameter representation, which is used to calculate the soft-photon corrections to the amplitude $f_C(q)$.

Unlike $f_C(q)$, the hadronic amplitude $f_N(q)$ is known only from experiment, and its Fourier transform \eqref{eq:Fourier} therefore yields the amplitude $\gamma_N(b)$. Since, in the Glauber approach, the calculation of a $pA$ amplitude relies on eikonal phases rather than amplitudes, it may be reasonable to make the following replacement in Eq.\,\eqref{eq:gNpA}:
\begin{equation}
  i\gamma_N^{pN}(b) \to i\chi_N^{pN}(b) = \int{
    d^2\boldsymbol{s}\ln\big(1+i\gamma_N(s)\big
    )
   \frac{T_A(\boldsymbol{b}\!-\!\boldsymbol{s})}{A}
  }
  \label{eq:chiNpC}
\end{equation}

\begin{figure*}[t]
  \begin{center}
    \includegraphics[width=0.70\columnwidth]{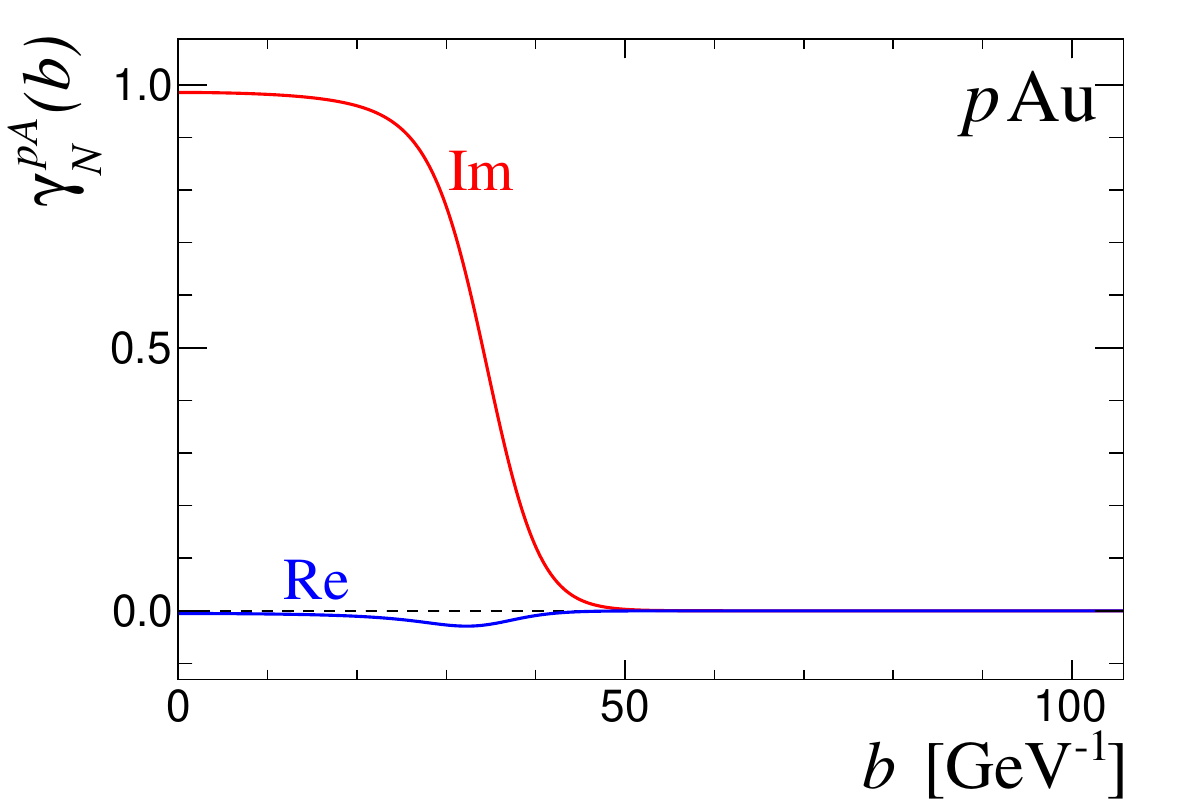}
    \hspace{2cm} 
    \includegraphics[width=0.70\columnwidth]{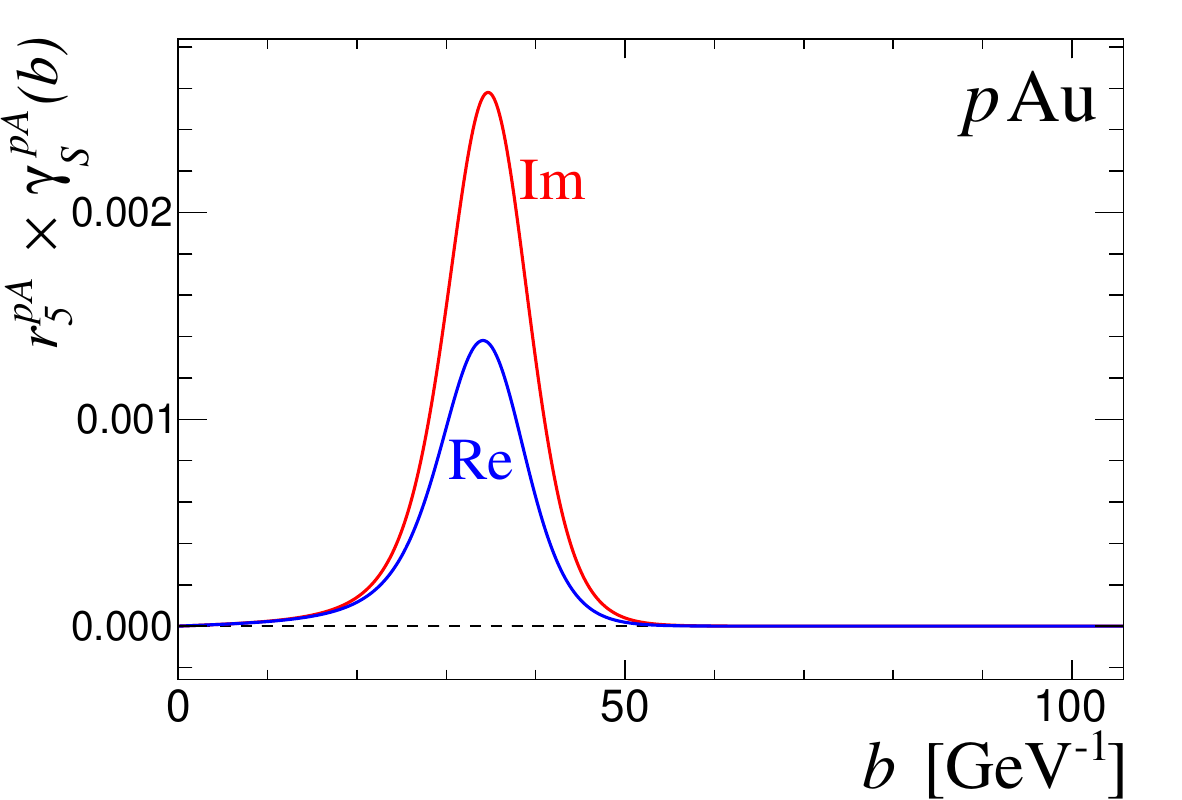}
  \end{center}
  \vspace{-1em}
  \caption{ \label{fig:fbNS}  
    Real and imaginary parts of the forward elastic $p$Au hadronic nonflip $\gamma_N^{p\text{Au}}(b)$ and spin-flip $r_5^{p\text{Au}}\gamma_S^{p\text{Au}}(b)$ amplitudes in impact-parameter space.
  }
\end{figure*}

However, the main goal of this paper is to provide a parametrization for calculating $A_\text{N}(t)$ with a small number of free parameters that can accurately approximate the measured analyzing power. We found that, for this purpose, better results can be obtained by using $\gamma_N^{pN}$ in Eq.\,\eqref{eq:gNpA}. Consequently, we will not apply the transformation in Eq.\,\eqref{eq:chiNpC} in this paper.

Similarly, following the approach used in Eq.\,\eqref{eq:TAh}, one can calculate the effective proton-nucleon eikonal amplitudes or phases $\gamma_S^{pN}(b)$, $\gamma_M^{pN}(b)$, and $\chi_C^{pN}(b)$. For exponential $pp$ form factors, the results for the spin-flip amplitudes $S$ and $M$ are the same as that for the nonflip hadronic amplitude in Eq.\,\eqref{eq:gNpN}. For the electromagnetic nonflip amplitude $C$, one obtains
\begin{equation}
  \chi_C^{pN}(b) = \chi_C(b,B_C+B_A) + \ln{\frac{B_C+B_A}{B_C}}
  \label{eq:chiCpN}
\end{equation}
Since the phase $\chi_C(b,B_C+B_A)$ is defined only up to an additive constant, the logarithmic term in Eq.\,\eqref{eq:chiCpN} can be omitted. We thus arrive at a conclusion similar to that expressed in Eq.\,\eqref{eq:gNpN}.

Because of the long-range nature of the electromagnetic interaction, the proton-nucleus amplitude can be written as
\begin{equation}
  \chi_C^{pA}(b) = Z\chi_C^{pN}(b) = Z\chi_C(b,B_C^{pA}) + \delta\chi_C^{pA}(b)
  \label{eq:chiCpA}
\end{equation}
where $Z$ is the nuclear charge and
\begin{equation}
  B_C^{pA} = B_C\,\exp{\left(\chi_C^{pN}(0)+\gamma\right)}.
  \label{eq:BCpA}
\end{equation}
As explained in Section\,\ref{sec:Coulomb}, separating the electromagnetic amplitudes into exponential and non-exponential parts is convenient for calculating the Coulomb corrections, i.e., the soft-photon corrections in the final states.

Similarly, using the same exponential form-factor slope $B_C^{pA}$, the proton-nucleus electromagnetic spin-flip amplitude can be decomposed as
\begin{equation}
  \gamma_M^{pA}(b) = Z\chi_M^{pN}(b) = Z\gamma_M(b,B_C^{pN}) + \delta\gamma_M^{pA}(b).
  \label{eq:gMpA}
\end{equation}

To evaluate multiple-scattering effects on the hadronic spin-flip amplitudes, the spin-flip and nonflip contributions from both $pp$ and $pn$ scattering should be considered simultaneously. This can be achieved by making the replacement
\begin{equation}
  i\gamma_N^{pN}(b) \to i\gamma_N^{pN}(b) + i r_5^{pN}\gamma_S^{pN}(b)
\end{equation}
in Eq.\,\eqref{eq:gNpA}. Since the factors $r_5^{pN}$ and $q/m_p$ entering the spin-flip amplitude are small, we retain only the leading-order contribution of the term $ir_5^{pN}\gamma_S^{pN}(b)$ to the proton-nucleus amplitude. This gives
\begin{equation}
  r_5^{pA}\times\gamma_S^{pA}(b) = r_5^{pN}\times A\gamma_S^{pN}(b) \left[ 1+i\gamma_N^{pN}(b)\right]^{A-1}.
  \label{eq:gSpA}
\end{equation}
If the parametrizations of the $pn$ and $pp$ spin-flip amplitudes, expressed in terms of $r_5$, are identical, then $r_5^{pN}=r_5^{pp}$. More generally, the effective value of $r_5^{pN}$ averaged over the protons and neutrons in the nucleus is given by
\begin{equation}
  r_5^{pN} = \frac{Zr_5^{pp} + (A-Z)r_5^{pn}}{A}.
  \label{eq:r5pA}
\end{equation}

In Ref.\,\cite{Kopeliovich:2023xtu}, it was emphasized that a beam energy of 100\,GeV in the laboratory frame is not sufficiently high to suppress isovector Reggeons, which have relatively large spin-flip components. However, for isoscalar nuclei, such as $^{12}$C and $^{16}$O, isovector Reggeons are completely excluded; for other nuclei, their contributions are suppressed by the small factor $(A-2Z)/A$ \cite{Kopeliovich:2023xtu}.

In other words, the difference between the $pp$ and $pn$ spin-flip amplitudes is important for the estimates presented here. The proton-proton parameter $r_5^{pp}$ was measured precisely at HJET \cite{Poblaguev:2019saw}. Although an accurate experimental value of $r_5^{pn}$ is not available at a beam energy of 100\,GeV, it can be evaluated using the results of the Regge fit to the HJET data \cite{Poblaguev:2019saw}. Details, together with the values of $r_5^{pA}$ for all nuclei considered in this paper, are provided in Appendix\,\ref{sec:r5}.

\begin{figure*}[t]
  \begin{center}
      \includegraphics[width=0.70\columnwidth]{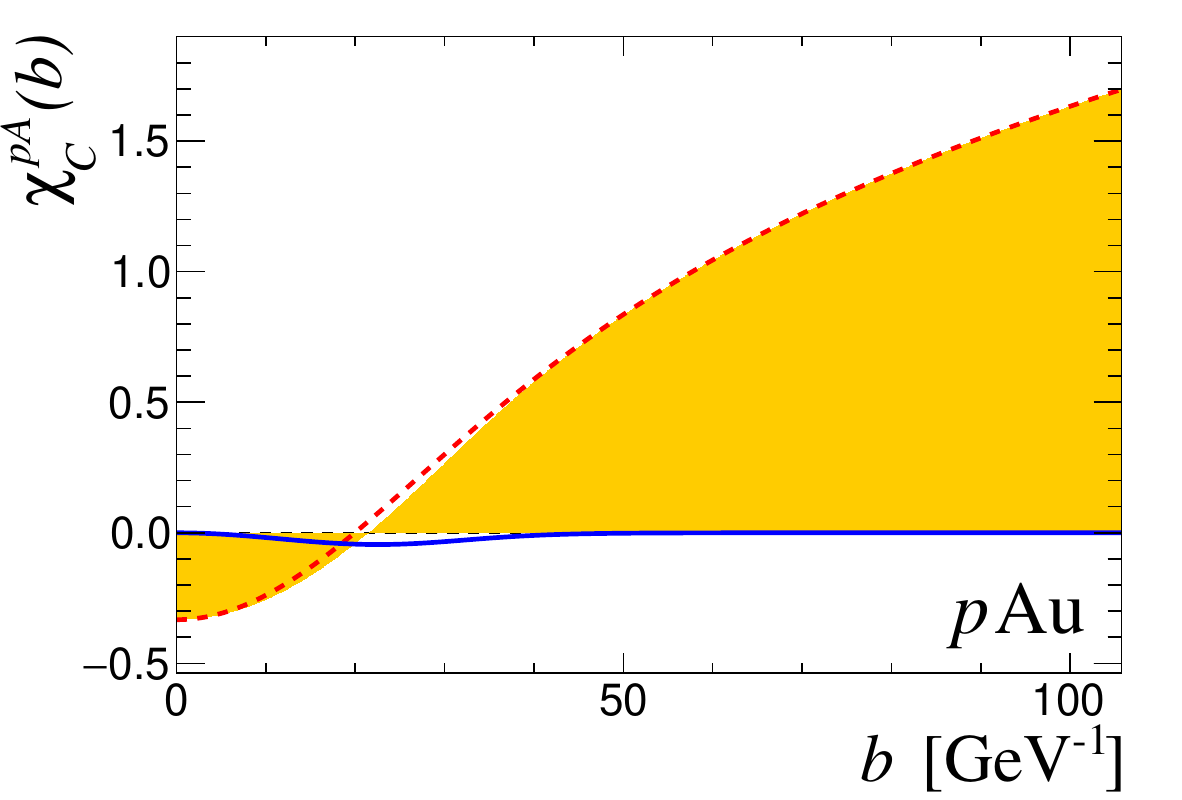}
      \hspace{2cm}
      \includegraphics[width=0.70\columnwidth]{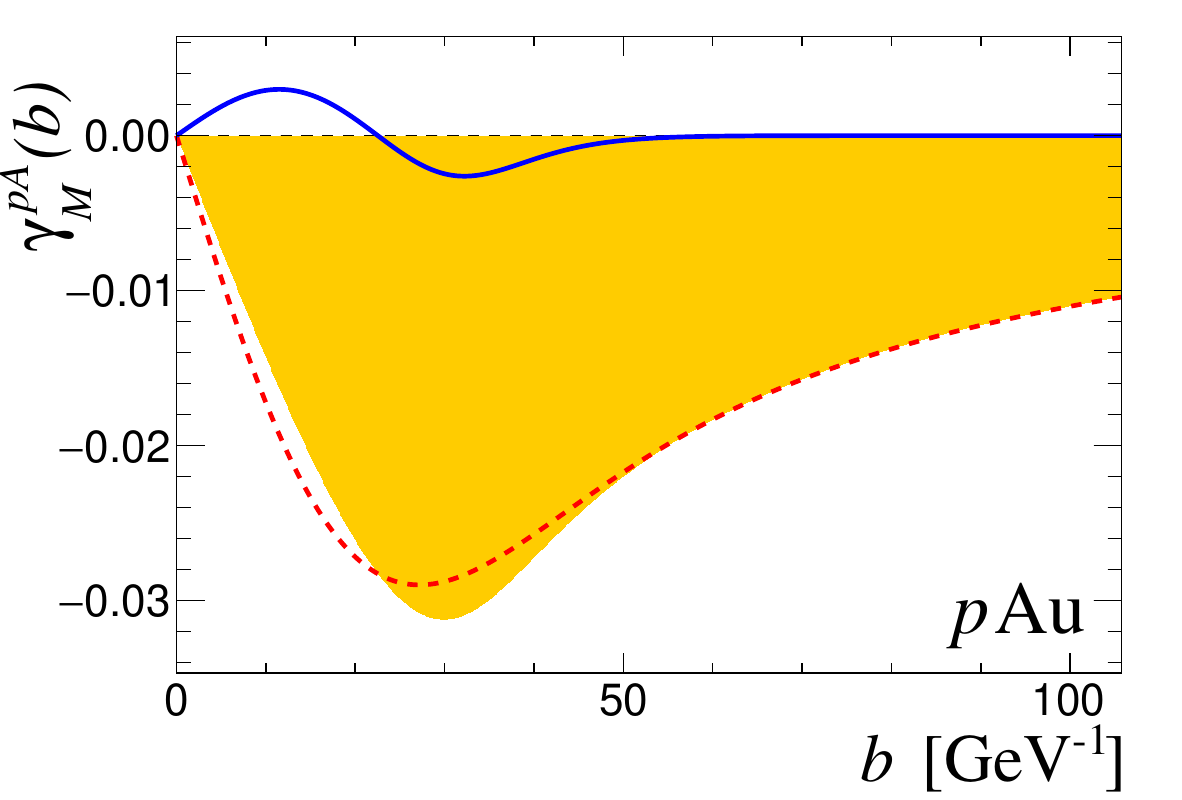}
    \end{center}
  \vspace{-1em}
  \caption{ \label{fig:fbCM}
    The eikonal phases for the electromagnetic amplitudes (filled areas). The exponential contributions are shown by dashed lines, and the non-exponential corrections by solid lines.
  }
\end{figure*}

\section{Elastic $\boldsymbol{p^\uparrow\!\!\!\text{Au}}$ scattering}

To evaluate the $p$Au amplitudes, we use the nuclear density parameters $R_A$ and $a$, and the hadronic spin-flip parameter $r_5^{pA}$ given in Tables \ref{tab:WS} and \ref{tab:r5A}, respectively.  

The hadronic $p^\uparrow\text{Au}$ amplitudes $\gamma_N^{p\text{Au}}(b)$ and $r_5^{p\text{Au}}\gamma_S^{p\text{Au}}(b)$ in impact-parameter space are shown in Fig.\,\ref{fig:fbNS}. Since these amplitudes are nonzero only over a limited range of $b$, the corresponding momentum-space amplitudes can be calculated by straightforward numerical integration:
\begin{align}
  f_N^{p\text{Au}}(q) &= \int_0^{b_\text{cut}}{db\,b\gamma_N^{p\text{Au}}(b) J_0(bq)}, \\
  f_S^{p\text{Au}}(q) &= r_5^{p\text{Au}}\int_0^{b_\text{cut}}{db\,b\gamma_S^{p\text{Au}}(b) J_1(bq)},
\end{align}
where $J_0(x)$ and $J_1(x)$ are Bessel functions. In the example considered here, we use $b_\text{cut}\approx105\,\text{GeV}^{-1}$. For practical calculations of the spin-flip amplitude, one can replace $i\boldsymbol{n\!\cdot\!b}$ with $b$ in the impact-parameter amplitude and then multiply the result by the angular factor $\boldsymbol{n\!\cdot\!q}/q$.

Numerical integration of the electromagnetic amplitudes (Fig.\,\ref{fig:fbCM}) requires a very large value of $b_\text{cut}$; moreover, the integral for the Coulomb amplitude (C) diverges. However, this problem can be readily overcome by separating the electromagnetic amplitudes into exponential and non-exponential parts:
\begin{align}
  f_C^{p\text{Au}}(q) &= \frac{-2\alpha Z}{q^2}e^{-B_C^{p\text{Au}} q^2/2} \nonumber \\
  &+ \int_0^{b_\text{cut}}{db\,b\,\delta\chi_C^{p\text{Au}}(b) J_0(bq)},
  \\
  f_M^{p\text{Au}}(q) &= -\frac{\kappa_p\alpha Z}{m_p} \frac{\boldsymbol{n\!\cdot\!q}}{q^2}e^{-B_C^{p\text{Au}} q^2/2} \nonumber \\
    &+ \int_0^{b_\text{cut}}{db\,b\,\delta\gamma_M^{p\text{Au}}(b) J_1(bq)}.
\end{align}

According to Eq.\,\eqref{eq:BCpA}, $B_C^{p\text{Au}}=294\,\text{GeV}^{-2}$.
The $p$Au nonflip amplitude can be written as
\begin{equation}
  f_\text{nf}^{p\text{Au}}(t) = \frac{\sigma_\text{tot}^{p\text{Au}}}{4\pi}\left[%
      i{\cal{F}}_\text{Im}(t) + \rho^{p\text{Au}}{\cal{F}}_\text{Re}(t) +\frac{t_c^{p\text{Au}}}{t}{\cal{F}}_C(t)
      \right].
  \label{eq:FF}
\end{equation}
Using the calculated value of $f_N^{p\text{Au}}(0)$, we obtain $\sigma_\text{tot}^{p\text{Au}}=2836\,\text{mb}$ and $-t_c^{p\text{Au}}=1.99\times10^{-3}\,\text{GeV}^2$.

\begin{figure}[t]
  \begin{center}
    \includegraphics[width=0.70\columnwidth]{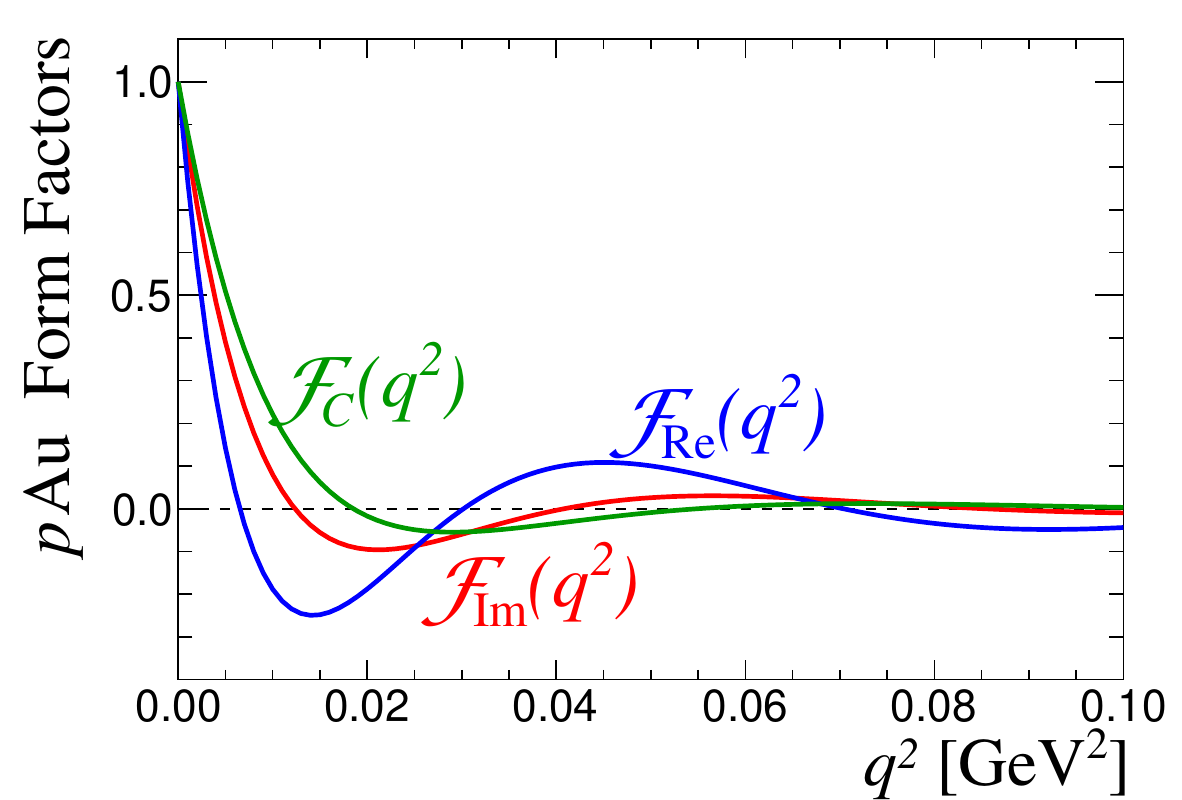}
  \end{center}
  \vspace{-1em}
  \caption{ \label{fig:FF}
    The $p$Au form factors, as defined in Eq.\,\eqref{eq:FF}.
  }
\end{figure}

The dependence of the form factors ${\cal F}(t)$ on the momentum transfer $t\approx -q^2$ is shown in Fig.\,\ref{fig:FF}.

\section{Coulomb Corrections \label{sec:Coulomb} }

\begin{figure}[t]
  \begin{center}
    \includegraphics[width=0.70\columnwidth]{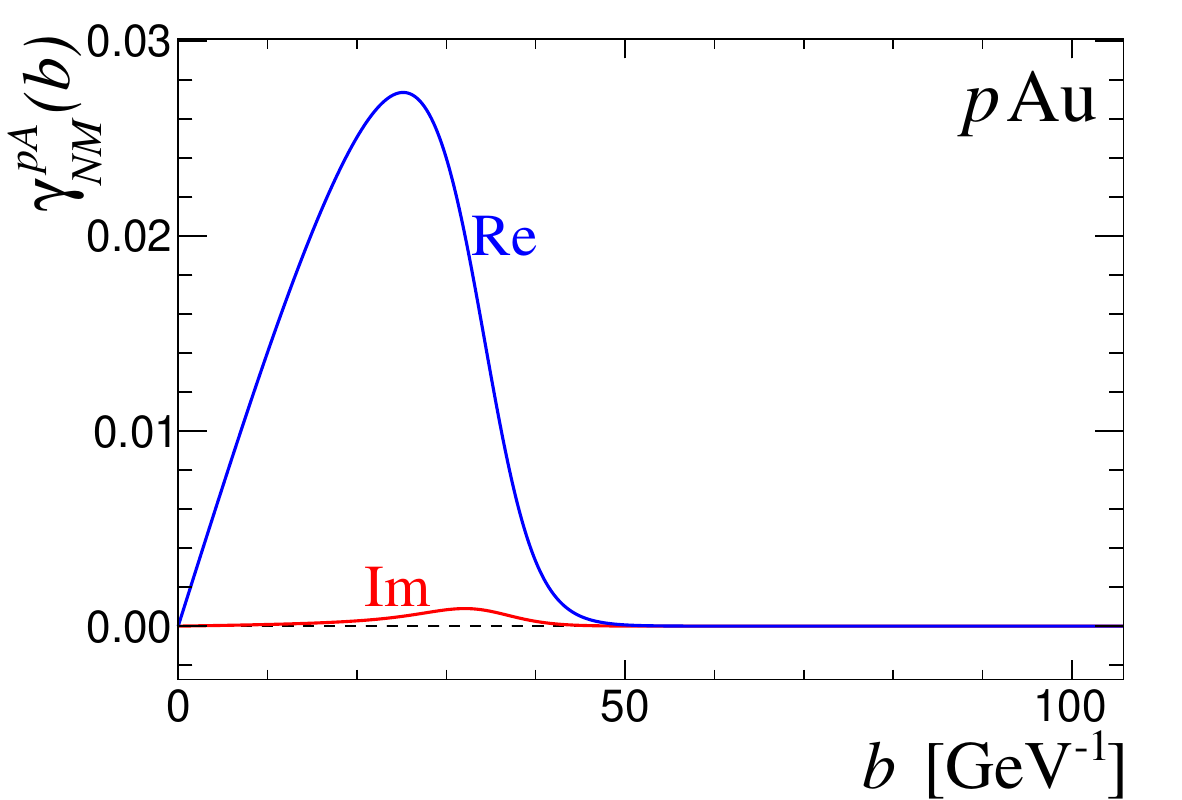}
  \end{center}
  \vspace{-1em}
  \caption{ \label{fig:fbNM}
    Real and imaginary parts of the effective spin-flip $p^\uparrow$Au amplitude arising from the magnetic-photon correction to the hadronic nonflip amplitude.
  }
\end{figure}

For elastic $pA$ scattering, one should account for modifications to the scattering amplitude,
\begin{equation}
  f^{pA}(\boldsymbol{q}) \to F^{pA}(\boldsymbol{q})
\end{equation}
due to multiple exchanges of soft photons. This effect can be treated using the eikonal approach. In general, $F^{pA}(\boldsymbol{q})$ can be expressed in terms of the profile function $i\Gamma(\boldsymbol{b})$ in impact-parameter space:
\begin{equation}
  F^{pA}(\boldsymbol{q})=\frac{1}{2\pi}\int{%
    d^2{\boldsymbol{b}}e^{-i\boldsymbol{q\cdot b}}\,i\Gamma(\boldsymbol{b}),
    \label{eq:invFourier}
  }
\end{equation}
where
\begin{equation}
  i\Gamma(\boldsymbol{b}) = i\left[1-e^{i\chi_N(\boldsymbol{b})+i\chi_S(b)+i\chi_C(b)+i\chi_M(\boldsymbol{b})}\right],
  \label{eq:Gamma}
\end{equation}
and the eikonal phases $\chi(b)$ are assumed to be obtained by Fourier transforming the corresponding amplitudes in the Born approximation. However, using Eqs.~\eqref{eq:chiCpA}, \eqref{eq:gNpA}, \eqref{eq:gSpA}, and~\eqref{eq:gMpA}, the expression \eqref{eq:Gamma} for the $pA$ profile function can be rewritten \cite{Poblaguev:2025vtd} as
\begin{align}
  i\Gamma(b) &= i\left[1-e^{i\chi_C^{pA}}\right] +\left[\gamma_N^{pA}+r_5^{pA}\gamma_S^{pA}+\gamma_M^{pA}\right] e^{i\chi_C^{pA}}
  \nonumber \\
  &+\left[i\gamma_N^{pA}\gamma_M^{pA}+ir_5^{pA}\gamma_S^{pA}\gamma_M^{pA}\right] e^{i\chi_C^{pA}}.
  \label{eq:GammaEff}
\end{align}
Notably, adding a constant term to $\chi_C^{pA}(b)$ changes only the common phase of all amplitudes contributing to $F^{pA}(\boldsymbol{q})$ and therefore does not affect any experimental observables. The eikonal functions $\gamma_N^{pA}(b)$, $\gamma_S^{pA}(b)$, and $\gamma_M^{pA}(b)$ in Eq.\,\eqref{eq:GammaEff} are similar to those defined in Eqs. \eqref{eq:gN}, \eqref{eq:gS}, and \eqref{eq:gM}, respectively, but without assuming an exponential form factor. Since the term $ir_5^{pA}\gamma_S^{pA}(b)\gamma_M^{pA}(b)$ provides only a small correction to $\gamma_N^{pA}(b)$, it is neglected.  The effective spin-flip amplitude
\begin{equation}
  \gamma_{N\!M}^{pA}(b) = i\gamma_N^{pA}(b)\gamma_M^{pA}(b)
\end{equation}
describes the correction to the hadronic nonflip amplitude arising from soft (ultraperipheral) magnetic-photon exchange. In elastic $p^\uparrow{p}$ scattering, this contribution produces a small but non-negligible correction to the measured $r_5$ \cite{Poblaguev:2021xkd}. In $p^\uparrow{A}$ scattering, however, the large nuclear charge $Z$ can make $\gamma_{N\!M}^{pA}(b)$ substantial (see Fig.\,\ref{fig:fbNM}), potentially leading to a significant modification of the squared-momentum-transfer dependence of the analyzing power $A_\text{N}^{pA}(t)$.

A well-known difficulty in evaluating Coulomb corrections is the divergence of the Fourier integral used to calculate $\chi_C(b)$. A fictitious photon mass is usually introduced to avoid this divergence. However, this approach substantially complicates the calculation of Coulomb corrections because it requires inverse Fourier integration over a wide range of impact parameters. It was noted in \cite{Poblaguev:2026wmb} that if the amplitudes $f_C^{pA}(q)$ and $f_M^{pA}(\boldsymbol{q})$ have the same exponential form factor $\exp{(-B_C^{pA}q^2/2)}$, their Coulomb corrections are identical:
\begin{equation}
  \frac{F_C^{pA}(q)}{f_C^{pA}(q)} = \frac{F_M^{pA}(\boldsymbol{q})}{f_M^{pA}(\boldsymbol{q})} = \widetilde{\mathcal{F}}\left(B_C^{pA}q^2/2\right),
\end{equation}
where, for fixed $Z$, $\widetilde{\mathcal{F}}(x)$ is a parameter-free function that can be calculated explicitly in the massless-photon limit \cite{Poblaguev:2025vtd} for the spin-flip amplitude. For completeness, the details of this calculation are provided in Appendix\,\ref{sec:CC}.

In the proposed approach,
\begin{equation}
  \chi_C^{pA}(b) = \chi_C^\text{exp}(b) + \delta\chi_C^{pA}(b),
\end{equation}
where
\begin{equation}
  \chi_C^\text{exp}(b) = \alpha Z 
  \left[\ln\left(\frac{b^2}{2B_C^{pA}}\right) + E_1\left(\frac{b^2}{2B_C^{pA}}\right)\right],  \label{eq:chiCF}
\end{equation}
is the contribution associated with the exponential form factor, and $\delta\chi_C^{pA}(b)$ is defined in Eq.\,\eqref{eq:chiCpA}. Similarly,
\begin{align}
  \gamma_M^{pA}(b) &= \gamma_M^\text{exp}(b) + \delta\gamma_M^{pA}(b), \\
  \gamma_M^\text{exp}(b) &= -\kappa_p\alpha{Z}\,i\frac{\boldsymbol{n\!\cdot\!b}}{m_p}\,\frac{1\!-\!e^{-b^2/2B_C^{pA}}}{b^2}, 
\end{align}

As a result, the Coulomb-corrected $pA$ amplitudes can be calculated over the limited integration range $0\text{\,--\,}b_\text{cut}$:
\begin{align}
  F_N^{pA}(q) &= \int_0^{b_\text{cut}}{ db\,b\,\gamma_Ne^{i\chi_C} J_0(bq)},
  \\
  F_S^{pA}(q) &= r_5^{pA}\times\int_0^{b_\text{cut}}{db\,b\,\gamma_Se^{i\chi_C}  J_1(bq)},
  \\
  F_{N\!M}^{pA}(q) &= \int_0^{b_\text{cut}}{db\,b\,\gamma_{N\!M}e^{i\chi_C} J_1(bq)},
  \\
  F_C^{pA}(q) &= \frac{-2\alpha Z}{q^2} e^{-B_Cq^2/2}\widetilde{\mathcal{F}}\left(B_Cq^2/2\right) \nonumber \\
  &+ \int_0^{b_\text{cut}}{ db\,b\,i\left[ 1-e^{i\delta\chi_C}\right]e^{i\chi_C^\text{exp}} J_0(bq) }, 
  \\
  F_M^{pA}(q) &= \frac{-\kappa_p\alpha Z}{m_pq}e^{-B_Cq^2/2}\widetilde{\mathcal{F}}\left(B_Cq^2/2\right) \nonumber \\
  &- \int_0^{b_\text{cut}}{db\,b\,\gamma_M\left[1-e^{i\delta\chi_C}\right]e^{i\chi_C^\text{exp}} J_1(bq)} \nonumber \\
  &+ \int_0^{b_\text{cut}}{db\,b\,\delta\gamma_Me^{i\chi_C} J_1(bq)}.
\end{align}
Here, for readability, $\chi_C\equiv\chi_C(b)$, $\gamma_N\equiv\gamma_N(b)$, etc.\@, and the superscripts ${}^{pA}$ are omitted on the right-hand sides of the equations.

\begin{figure}[t]
  \begin{center}
    \includegraphics[width=0.70\columnwidth]{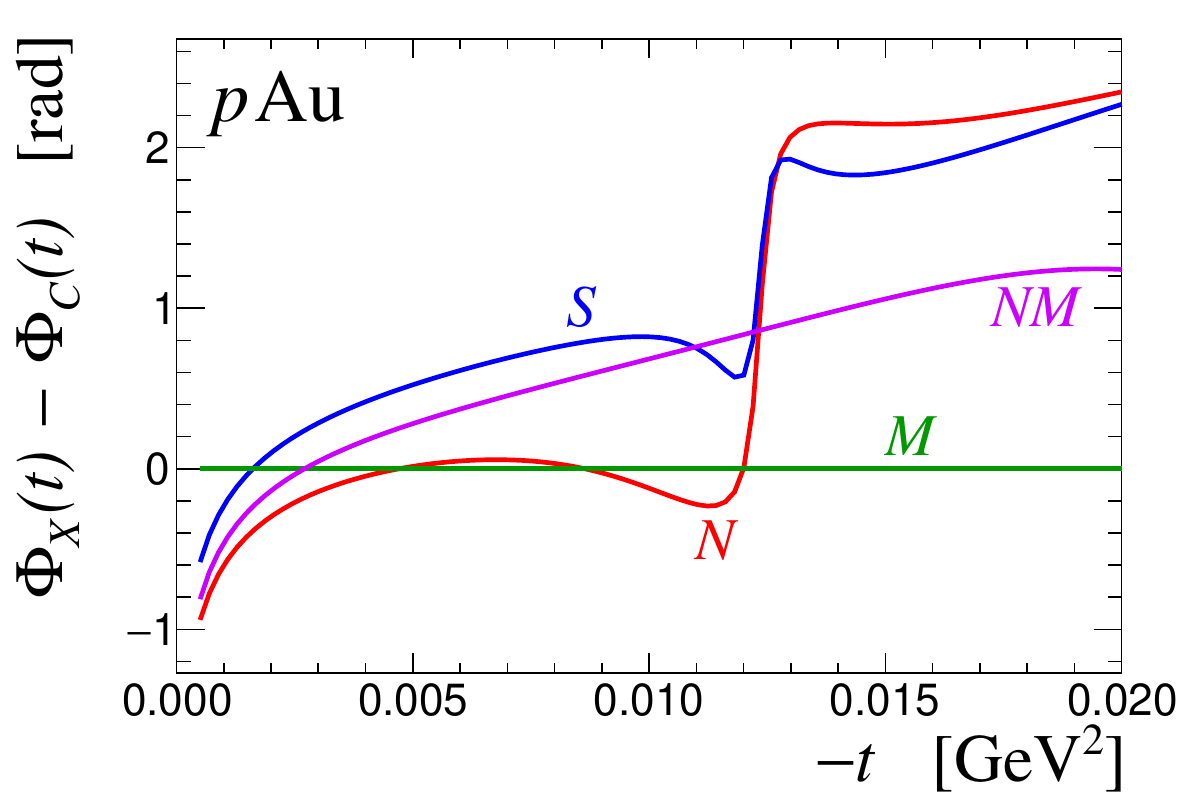}
  \end{center}
  \vspace{-1em}
  \caption{ \label{fig:Theta}
    The calculated Coulomb phases $\Phi_X(t)$, where $X=\{N, S, M, N\!M\}$, compared with $\Phi_C(t)$.
  }
\end{figure}

\begin{figure}[t]
  \begin{center}
    \includegraphics[width=0.70\columnwidth]{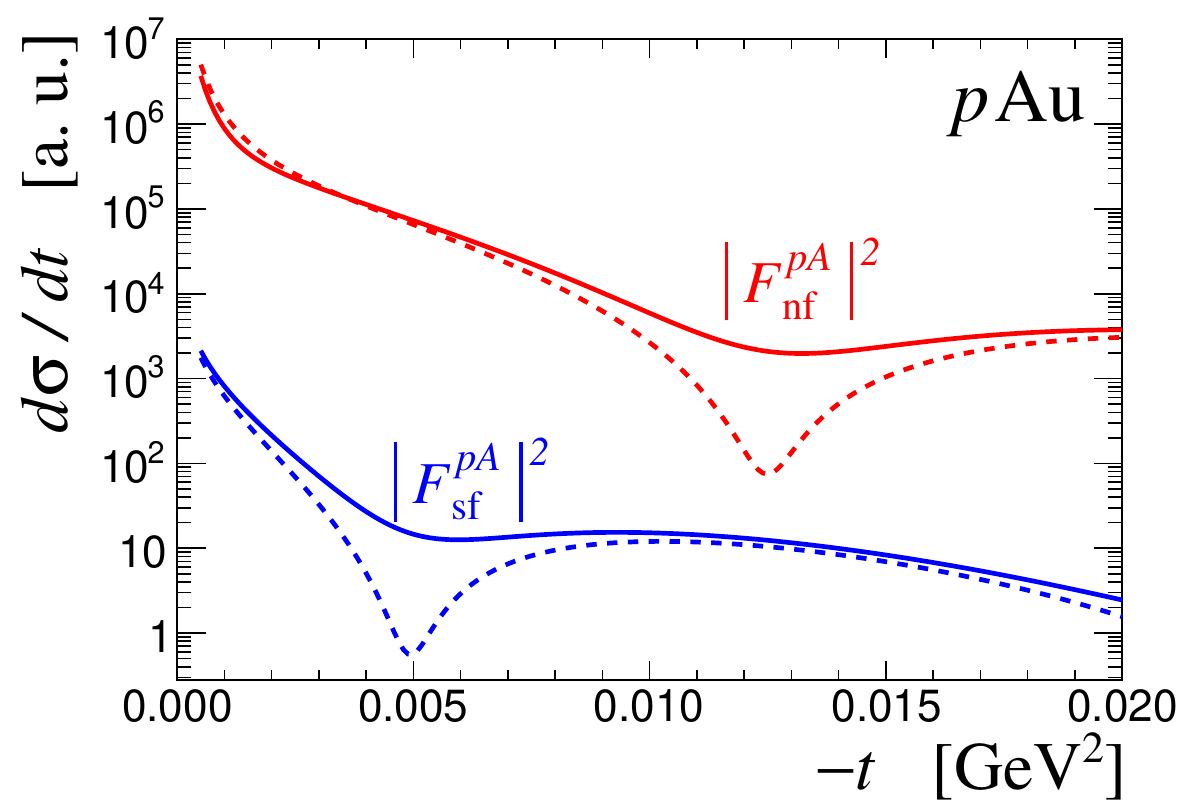}
  \end{center}
  \vspace{-1em}
  \caption{ \label{fig:cs}
    The calculated nonflip and spin-flip contributions to the differential $p^\uparrow$Au cross section. Dashed and solid lines show $d\sigma/dt$ before and after Coulomb corrections, respectively.
  }
\end{figure}

\begin{figure}[t]
  \begin{center}
    \includegraphics[width=0.90\columnwidth]{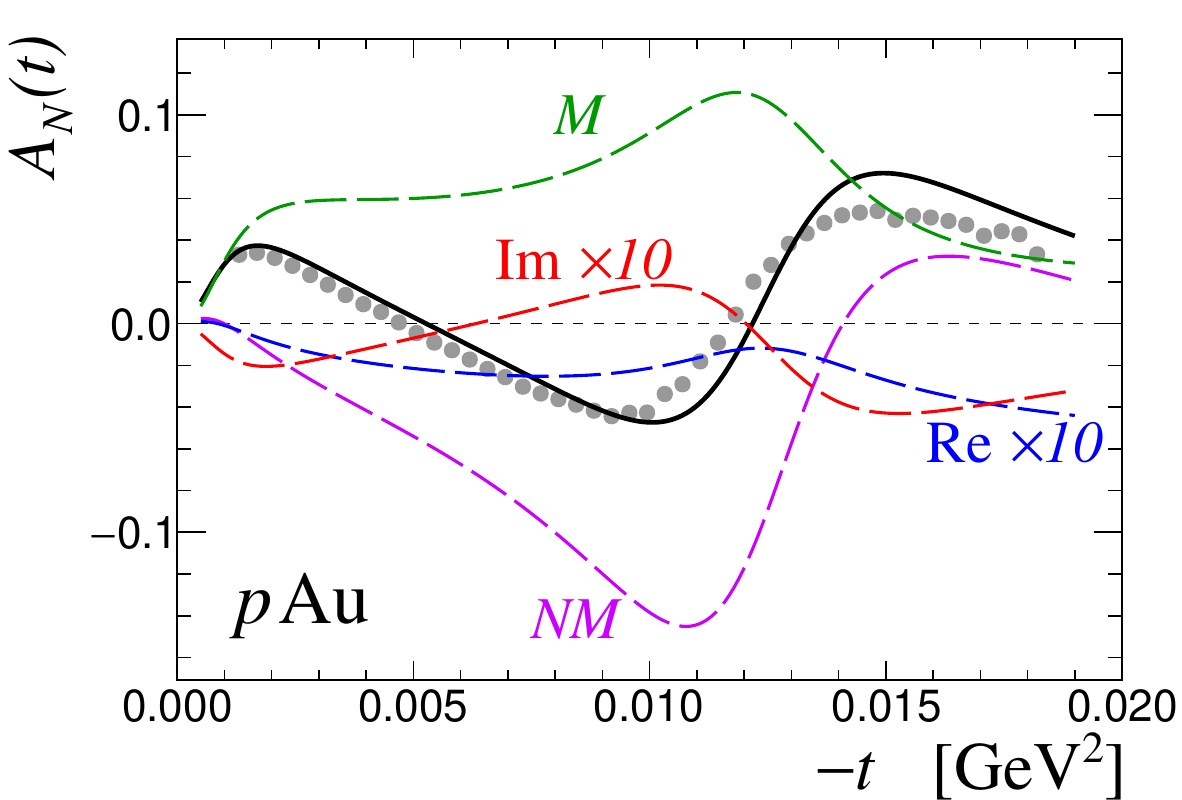}
  \end{center}
  \vspace{-1em}
  \caption{ \label{fig:Au_}
    The calculated $A_\text{N}^{p\text{Au}}(t)$ using the \emph{nominal} parametrization (solid line), compared with the measured values (solid circles).
    Dashed lines show the contributions from the four spin-flip amplitude components \eqref{eq:Fsf_exp}. The contributions associated with $r_5^\text{p\text{Au}}$ are scaled by a factor of 10.
  }
\end{figure}

The Coulomb phases, defined, for example, as
\begin{equation}
\Phi_N(t) = \arg\left[F_N^{pA}(t)/f_N^{pA}(t)\right],
\end{equation}
and measured relative to $\Phi_C(q)$, are shown for all $p^\uparrow$Au amplitudes in Fig.\,\ref{fig:Theta}. Notably, $\Phi_M(q)-\Phi_C(q)\approx0$, indicating that the non-exponential parts of the $M$ and $C$ form factors have a negligible effect on this phase difference. For the other amplitudes, the phase differences can vary substantially over the $t$-range covered by the HJET measurements. In particular, the phases of the $N$ and $S$ amplitudes exhibit sharp variations near the diffraction dip.

Coulomb corrections may also modify the absolute values of the amplitudes, as illustrated in Fig.\,\ref{fig:cs}.

\section{$\boldsymbol{p^\uparrow\!\!\!\text{Au}}$ analyzing power
\label{sec:pAu} }

\begingroup
\setlength{\tabcolsep}{10pt}
\begin{table*}[t]
  \caption{ \label{tab:Au} 
     Results for the $p^\uparrow$Au analyzing power fits. Parameters fixed in the fit are displayed by bold font.
  }
  \smallskip
  \begin{tabular}{ c | r r r r | r}
   \rule[-6pt]{0.pt}{0.5pt} $t$-Range [GeV$^2$] 
    & \multicolumn{1}{|c}{$R_A\;\;\text{[fm]}$}
    & \multicolumn{1}{ c}{$a\;\:\text{[fm]}$}
    & \multicolumn{1}{ c}{Re\,$r_5^{pA}$}
    & \multicolumn{1}{ c}{Im\,$r_5^{pA}$}
    & \multicolumn{1}{|c}{$\chi^2/\text{ndf}$}
    \\ \hline
    \multirow{8}{*}{0.001 -- 0.019}
    & \multicolumn{1}{c}{\bf 6.409} & \multicolumn{1}{c}{\bf 0.475} & \multicolumn{1}{c}{\bf 0.0176} & \multicolumn{1}{c|}{\bf 0.0427} & 10006.2\,/\,46 \\
    & \multicolumn{1}{c}{\bf 6.409} & \multicolumn{1}{c}{\bf 0.475} & $0.0593\pm0.0008$              & $0.0908\pm0.0016$               &  1355.1\,/\,44 \\
    & $6.547\pm0.019$               & $0.645\pm0.009$               & \multicolumn{1}{c}{\bf 0.0176} & \multicolumn{1}{c|}{\bf 0.0427} &  2876.9\,/\,44 \\
    & \multicolumn{1}{c}{\bf 6.409} & $0.603\pm0.004$               & $0.0523\pm0.0008$              & $0.0631\pm0.0018$               &   184.8\,/\,43 \\
    & $6.629\pm0.007$               & \multicolumn{1}{c}{\bf 0.475} & $0.0471\pm0.0009$              & $0.0800\pm0.0016$               &   439.7\,/\,43 \\
    & $6.801\pm0.019$               & $0.429\pm0.013$               & \multicolumn{1}{c}{\bf 0.0176} & $0.1069\pm0.0017$               &  1479.6\,/\,43 \\
    & $6.036\pm0.025$               & $0.755\pm0.009$               & $0.0620\pm0.0008$              & \multicolumn{1}{c|}{\bf 0.0427} &    50.1\,/\,43 \\
    & $6.017\pm0.037$               & $0.763\pm0.014$               & $0.0627\pm0.0013$              & $0.0407\pm0.0028$               &    49.6\,/\,42 \\
    \hline
    \multirow{8}{*}{0.001 -- 0.010}
    & \multicolumn{1}{c}{\bf 6.409} & \multicolumn{1}{c}{\bf 0.475} & \multicolumn{1}{c}{\bf 0.0176} & \multicolumn{1}{c|}{\bf 0.0427} &  8691.5\,/\,24 \\
    & \multicolumn{1}{c}{\bf 6.409} & \multicolumn{1}{c}{\bf 0.475} & $0.0631\pm0.0008$              & $0.0834\pm0.0016$               &   204.2\,/\,22 \\
    & $7.132\pm0.021$               & $0.436\pm0.013$               & \multicolumn{1}{c}{\bf 0.0176} & \multicolumn{1}{c|}{\bf 0.0427} &   385.8\,/\,22 \\
    & \multicolumn{1}{c}{\bf 6.409} & $0.628\pm0.009$               & $0.0505\pm0.0010$              & $0.0551\pm0.0025$               &    26.2\,/\,21 \\
    & $6.807\pm0.023$               & \multicolumn{1}{c}{\bf 0.475} & $0.0359\pm0.0017$              & $0.0674\pm0.0020$               &    47.7\,/\,21 \\
    & $7.198\pm0.022$               & $0.294\pm0.022$               & \multicolumn{1}{c}{\bf 0.0176} & $0.0776\pm0.0021$               &   107.6\,/\,21 \\
    & $5.954\pm0.095$               & $0.762\pm0.020$               & $0.0666\pm0.0036$              & \multicolumn{1}{c|}{\bf 0.0427} &    22.5\,/\,21 \\
    & $6.054\pm0.203$               & $0.733\pm0.057$               & $0.0633\pm0.0071$              & $0.0459\pm0.0059$               &    22.2\,/\,20 \\
  \end{tabular}
\end{table*}
\endgroup

Using the \emph{nominal values} of the nuclear density $\rho_A(r)$ (Table\,\ref{tab:WS}) and the hadronic spin-flip parameter $r_5^{pA}$ (Table\,\ref{tab:r5A}), one can calculate the Coulomb-corrected values of all five scattering amplitudes and subsequently determine the analyzing power $A_\text{N}^{pA}(t)$ for a given squared momentum transfer, $t=-q^2$. The results of these calculations for $p^\uparrow$Au scattering are compared with the HJET measurements in Fig.\,\ref{fig:Au_}. Given the complexity of deriving the $pA$ amplitudes, including Coulomb corrections, from the corresponding $pp$ amplitudes, the agreement between the calculated and measured $A_\text{N}^{p\text{Au}}(t)$ can be considered reasonably good. However, this agreement is insufficient for a precise determination of the proton-proton parameter $r_5$ from these measurements.

Expanding the spin-flip amplitude as
\begin{align}
  F_\text{sf}^{p\text{Au}}(\boldsymbol{q}) &=
  F_M^{p\text{Au}}(\boldsymbol{q}) + F_{N\!M}^{p\text{Au}}(\boldsymbol{q}) \nonumber
  \\ &+
  \operatorname{Re}r_5^{p\text{Au}} \frac{F_S^{p\text{Au}}(\boldsymbol{q})}{r_5^{p\text{Au}}} +
  i\,\operatorname{Im}r_5^{p\text{Au}} \frac{F_S^{p\text{Au}}(\boldsymbol{q})}{r_5^{p\text{Au}}},
  \label{eq:Fsf_exp}
\end{align}
and retaining only one term from the expansion \eqref{eq:Fsf_exp} in the numerator on the right-hand side of Eq.\,\eqref{eq:AN}, one can explicitly evaluate that term's contribution to the analyzing power. The results of these calculations for $p^\uparrow$Au scattering are presented in Fig.\,\ref{fig:Au_}.

The dominant contributions to $A_\text{N}^{p\text{Au}}(t)$ arise from the two spin-flip amplitudes $F_{M}^{p\text{Au}}(\boldsymbol{q})$ and $F_{N\!M}^{p\text{Au}}(\boldsymbol{q})$. Since the latter can be interpreted \cite{Kopeliovich:2023xtu} as an absorptive correction to $F_{M}^{p\text{Au}}(\boldsymbol{q})$, we arrive at the following conclusion: in high-energy, forward elastic $pA$ scattering, the analyzing power is governed primarily by the interaction of the proton's magnetic moment with the nuclear charge. However, the absorptive correction to this amplitude increases with the nuclear charge $Z$ and becomes substantial for heavy nuclei.

In Fig.\,\ref{fig:Au_}, the contributions from the hadronic spin-flip amplitude associated with $\operatorname{Re}r_5^{p\text{Au}}$ and $\operatorname{Im}r_5^{p\text{Au}}$ are shown separately. The corresponding curves are multiplied by a factor of 10 for better visibility.

% Keep the figure environment unchanged except for its caption, if desired:
\begin{figure}[b]
  \begin{center}
    \includegraphics[width=0.90\columnwidth]{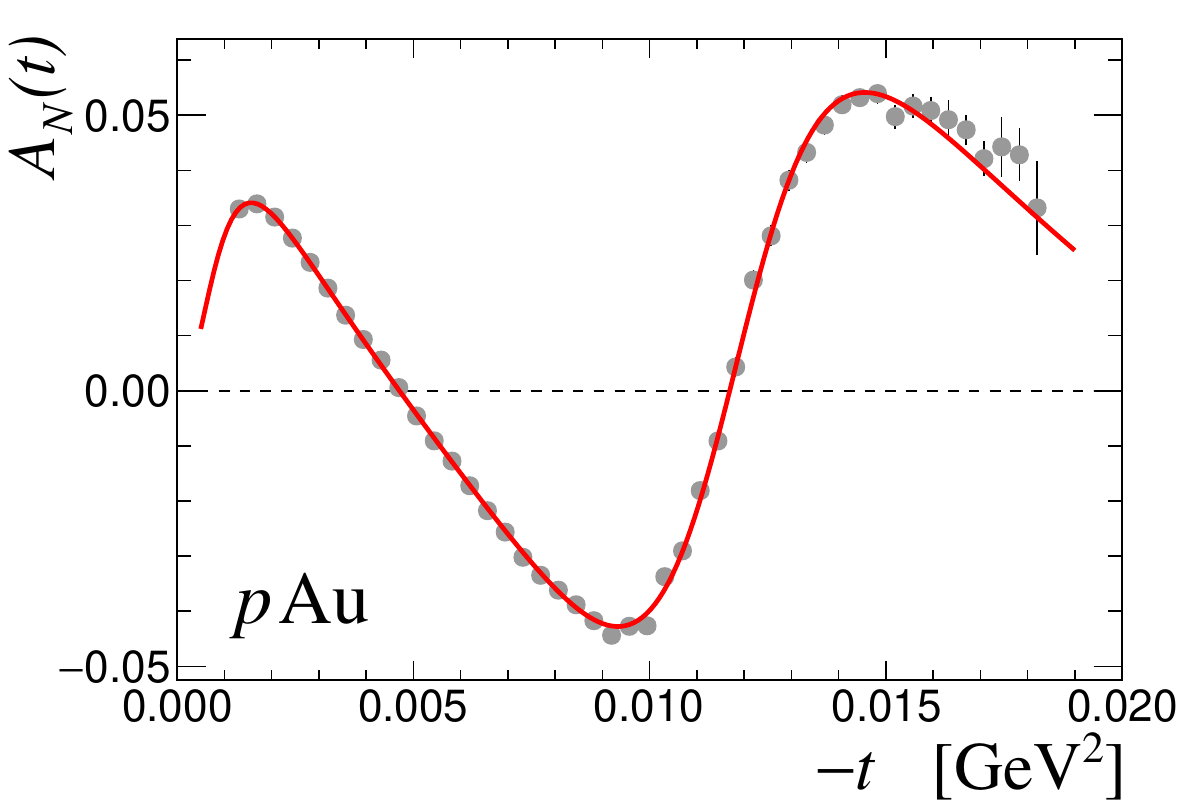}
  \end{center}
  \vspace{-1em}
  \caption{ \label{fig:Au}
    The fitted $A_\text{N}^{p\text{Au}}(t)$ (solid line) compared with the measured values (solid circles).
  }
\end{figure}

% Keep the figure environment and its placement unchanged:
\begin{figure*}[t]
  \begin{center}
    \includegraphics[width=0.70\columnwidth]{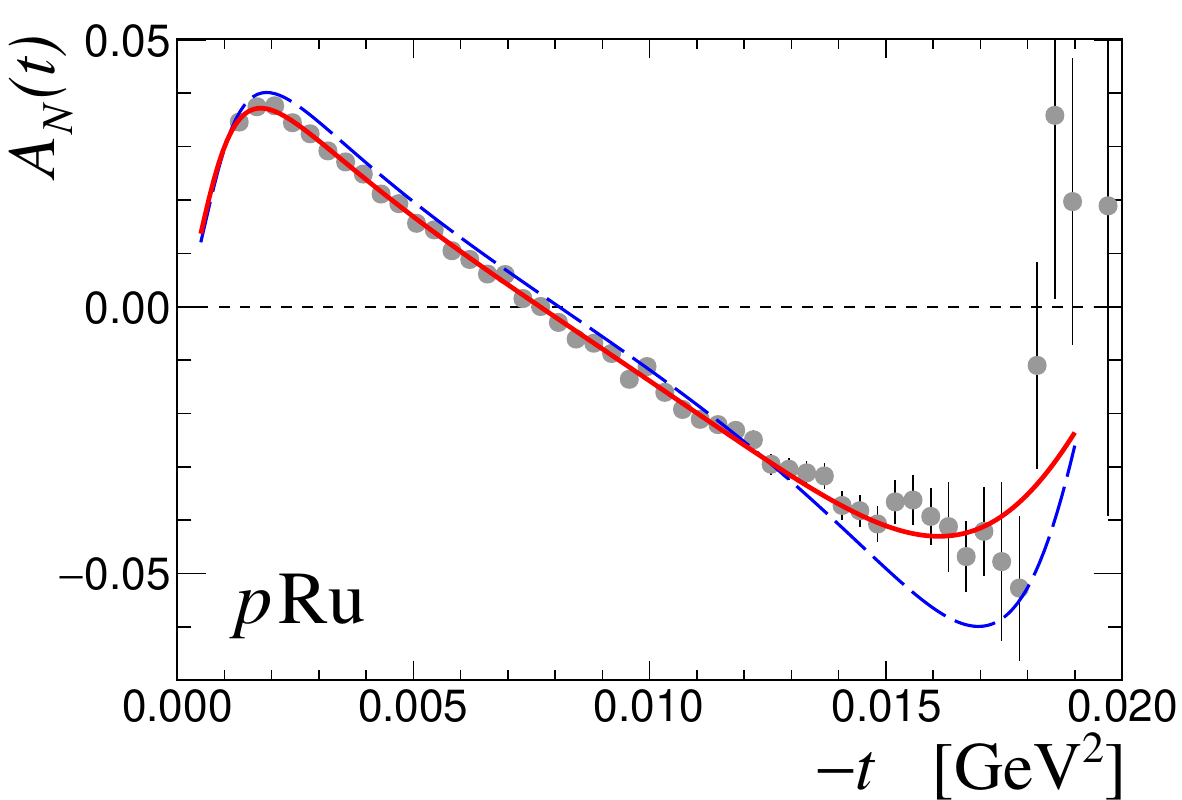}
    \hspace{1cm}
    \includegraphics[width=0.70\columnwidth]{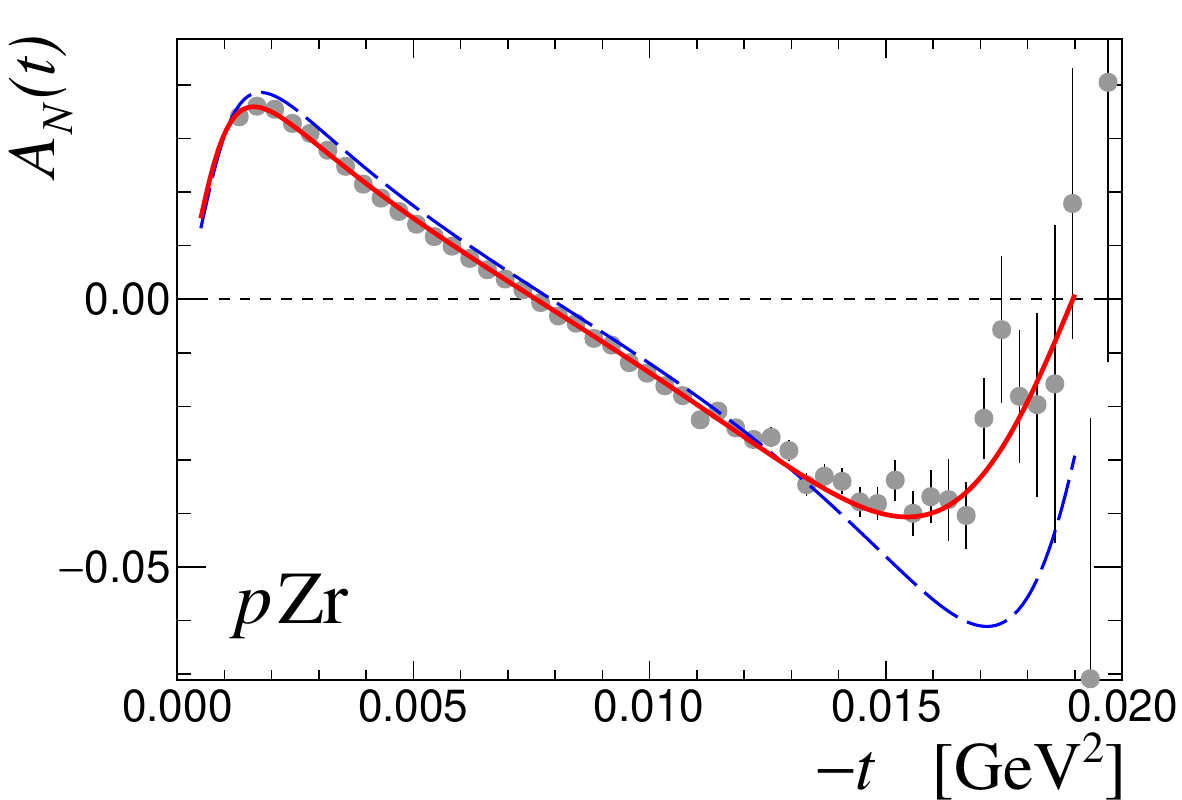}
    \includegraphics[width=0.70\columnwidth]{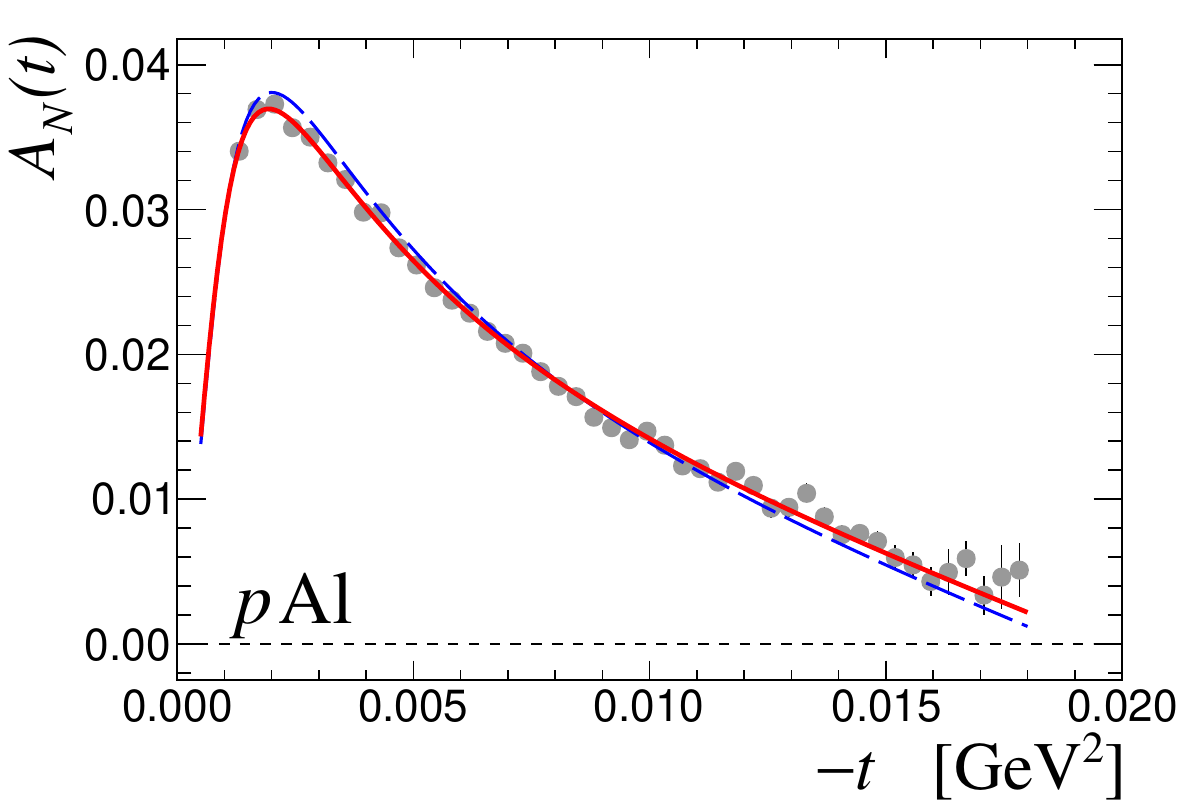}
    \hspace{1cm}
    \includegraphics[width=0.70\columnwidth]{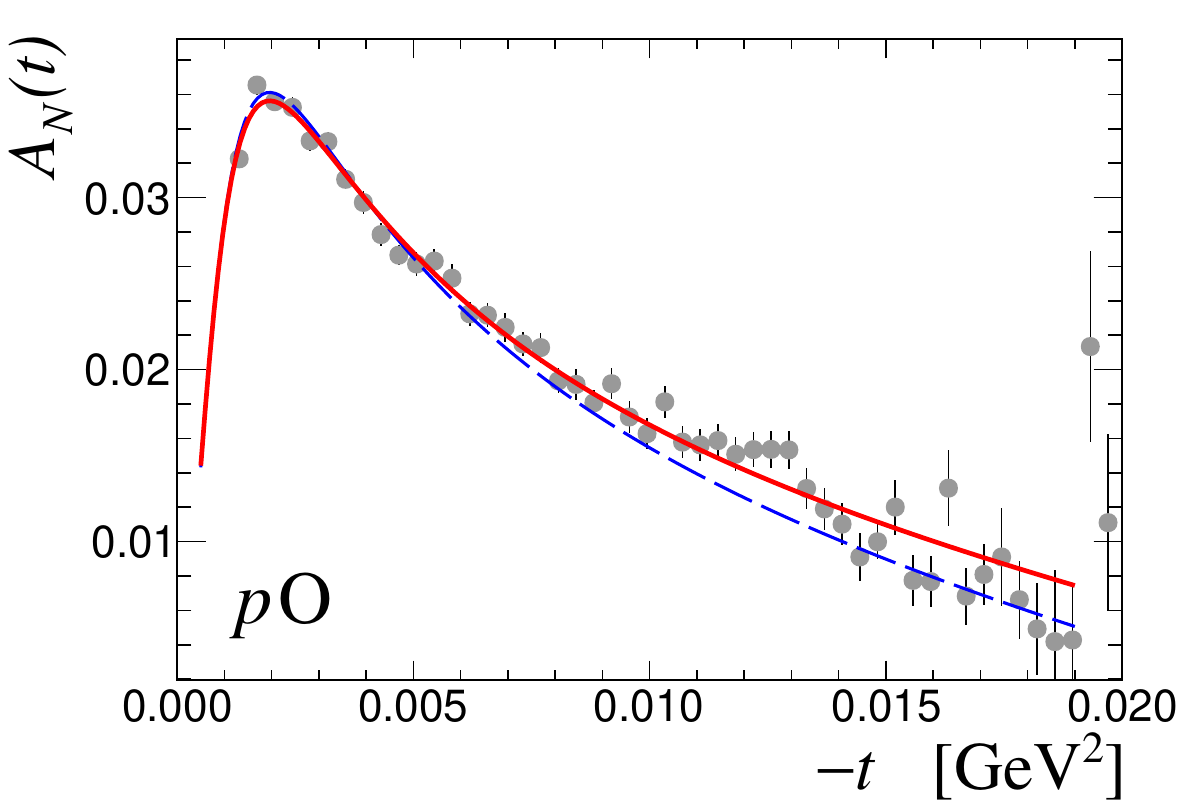}
    \includegraphics[width=0.70\columnwidth]{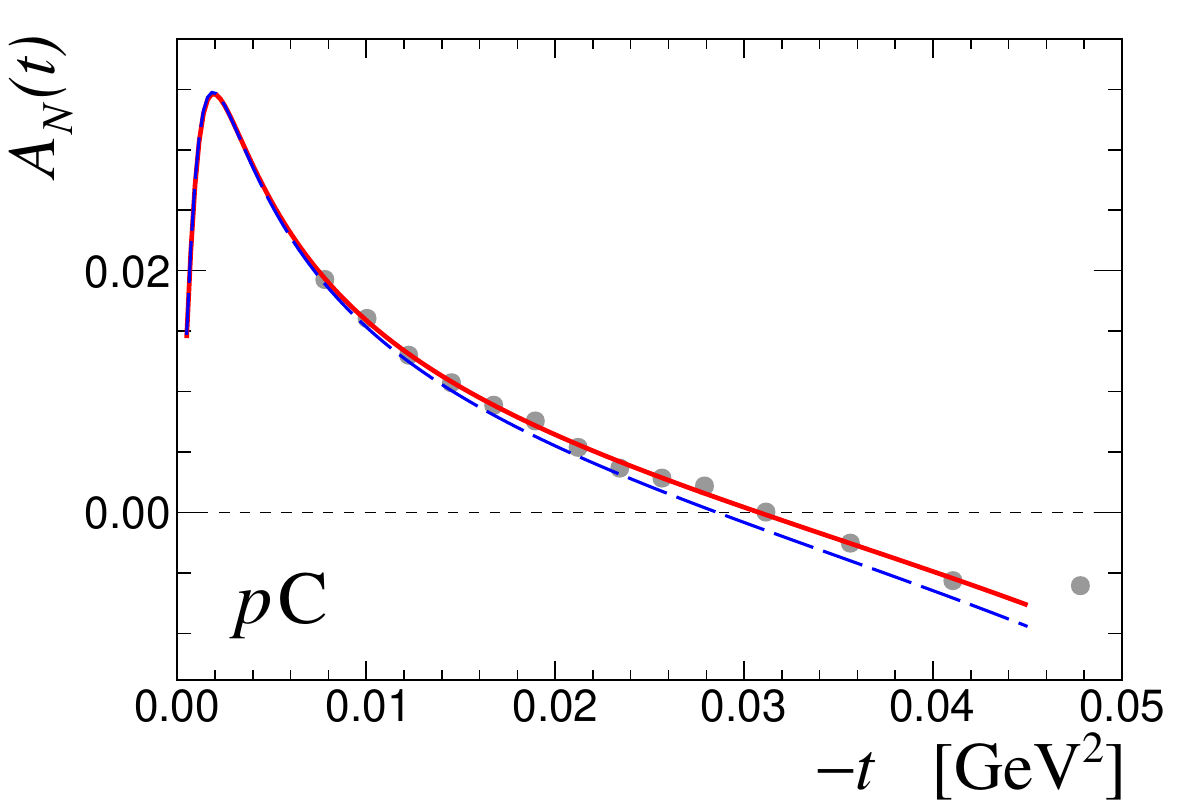}
    \hspace{1cm}
    \includegraphics[width=0.70\columnwidth]{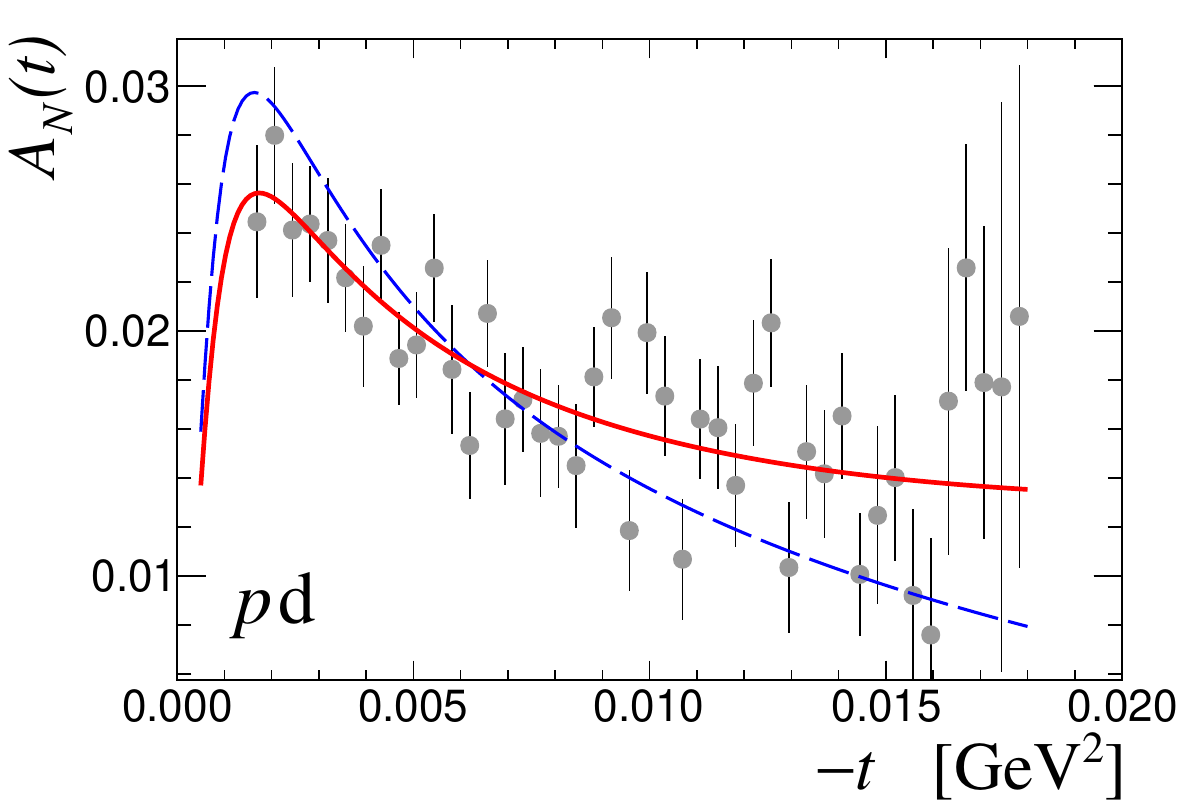}
  \end{center}
  \vspace{-1em}
  \caption{ \label{fig:Ru+}
    Analyzing powers for Ru, Zr, Al, O, C, and d. Solid circles represent the experimental data, dashed lines show calculations using the \emph{nominal values} of the parameters, and solid lines show the results of fits with free parameters.
  }
\end{figure*}

To improve the agreement between the data and the calculations, some or all of the parameters $R_A$, $a$, and the complex parameter $r_5$ were allowed to vary in the fit. The results are summarized in Table\,\ref{tab:Au}. Over the full range, $0.001<-t<0.019\,\text{GeV}^2$, the fit with all four parameters free (see Fig.\,\ref{fig:Au}) yields a $\chi^2$ value consistent with the number of degrees of freedom (ndf) at the 0.7-standard-deviation level.

It should be emphasized that fixing $\operatorname{Im}r_5^{p\text{Au}}=0.0427$ at its \emph{nominal value} does not reduce the quality of the fit. Fixing any of the other parameters at its nominal value, however, results in a substantial increase in $\chi^2$.

As shown in Fig.\,\ref{fig:Au}, a systematic discrepancy between the data and the calculations is observed for $|t|>0.015\,\text{GeV}^2$. This discrepancy may indicate an effect not accounted for in the model used for the fit. To investigate this possibility, the dependence of the fit results on the constraints imposed on the parameters $R_A$, $a$, and $r_5$ was also studied over a reduced $t$-range, $-t<0.010\,\text{GeV}^2$. As shown in Table\,\ref{tab:Au}, the fitted values of all parameters agree well between the two $t$-ranges.

Nevertheless, the possibility that the fit does not account for some additional effect cannot be excluded. For example, using the parameters obtained from the fit over the range $-t<0.010\,\text{GeV}^2$ to calculate $A_\text{N}^{p\text{Au}}(t)$ over the range $-t>0.010\,\text{GeV}^2$ gives
\begin{equation}
  \chi^2/\text{ndf} = 51.2/22,
\end{equation}
indicating that a single set of fitted values for $R_A$, $a$, and $r_5$ does not describe both $t$-ranges equally well.

Thus, some effect contributing to $A_\text{N}^{p\text{Au}}(t)$ is likely missing from the model used to describe the $p$Au analyzing power. However, the unaccounted-for corrections to $A_\text{N}^{p\text{Au}}(t)$ may be reasonably well approximated by variations in $R_A$, $a$, and $\operatorname{Re}r_5^{p\text{Au}}$. If this is the case, the corresponding parameters obtained from the analyzing-power fit should be interpreted as effective fit parameters rather than as the Woods-Saxon half-density radius and surface diffuseness parameter of gold or the real part of the hadronic spin-flip parameter $r_5^{p\text{Au}}$.

Potential sources of the unaccounted-for effects in $A_\text{N}^{p\text{Au}}(t)$ include, for example, systematic uncertainties in the HJET measurements, which have not yet been fully evaluated. Within the Glauber approach \cite{Glauber:1970jm}, additional corrections associated with center-of-mass correlations and nuclear breakup could also be considered. Furthermore, Ref.\,\cite{Kopeliovich:2023xtu} suggested considering Gribov inelastic shadowing corrections \cite{Gribov:1968jf,Kopeliovich:2016jjx}.

\begingroup
\setlength{\tabcolsep}{10pt}
\begin{table*}[t]
  \caption{ \label{tab:Ru}
     Analyzing power fit results. Parameters fixed in the fit are displayed by bold font.
  }
  \smallskip
  \begin{tabular}{ c | r r r r | r}
    \rule[-6pt]{0.pt}{0.5pt}  A 
    & \multicolumn{1}{|c}{$R_A\;\;\text{[fm]}$}
    & \multicolumn{1}{ c}{$a\;\:\text{[fm]}$}
    & \multicolumn{1}{ c}{Re\,$r_5^{pA}$}
    & \multicolumn{1}{ c}{Im\,$r_5^{pA}$}
    & \multicolumn{1}{|c}{$\chi^2/\text{ndf}$}
    \\
        \hline
    \multirow{4}{*}{ ${}^{197}$Au }
    & \multicolumn{1}{c}{\bf 6.409} & \multicolumn{1}{c}{\bf 0.475} & \multicolumn{1}{c}{\bf 0.0176} & \multicolumn{1}{c|}{\bf 0.0427} & 10006.2\,/\,46 \\
    & $6.547\pm0.019$               & $0.645\pm0.009$               & \multicolumn{1}{c}{\bf 0.0176} & \multicolumn{1}{c|}{\bf 0.0427} &  2876.9\,/\,44 \\
    & $6.036\pm0.025$               & $0.755\pm0.009$               & $0.0620\pm0.0008$              & \multicolumn{1}{c|}{\bf 0.0427} &    50.1\,/\,43 \\
    & $6.017\pm0.037$               & $0.763\pm0.014$               & $0.0627\pm0.0013$              & $0.0407\pm0.0028$               &    49.6\,/\,42 \\
    \hline
    \multirow{4}{*}{ ${}^{96}$Ru }
    & \multicolumn{1}{c}{\bf 4.995} & \multicolumn{1}{c}{\bf 0.507} & \multicolumn{1}{c}{\bf 0.0144} & \multicolumn{1}{c|}{\bf 0.0381} &   818.8\,/\,48 \\
    & $4.652\pm0.064$               & $0.857\pm0.023$               & \multicolumn{1}{c}{\bf 0.0144} & \multicolumn{1}{c|}{\bf 0.0381} &   112.7\,/\,46 \\
    & $3.816\pm0.226$               & $0.981\pm0.034$               & $0.0330\pm0.0039$              & \multicolumn{1}{c|}{\bf 0.0381} &    64.7\,/\,43 \\
    & $2.802\pm0.609$               & $1.215\pm0.099$               & $0.0466\pm0.0076$              & $-0.0004\pm0.0157$              &    56.9\,/\,42 \\
    \hline
    \multirow{4}{*}{ ${}^{96}$Zr }
    & \multicolumn{1}{c}{\bf 4.995} & \multicolumn{1}{c}{\bf 0.507} & \multicolumn{1}{c}{\bf 0.0167} & \multicolumn{1}{c|}{\bf 0.0415} &   813.9\,/\,48 \\
    & $4.559\pm0.067$               & $0.893\pm0.023$               & \multicolumn{1}{c}{\bf 0.0167} & \multicolumn{1}{c|}{\bf 0.0415} &    81.8\,/\,46 \\
    & $4.188\pm0.164$               & $0.951\pm0.031$               & $0.0244\pm0.0027$              & \multicolumn{1}{c|}{\bf 0.0415} &    69.6\,/\,43 \\
    & $3.549\pm0.321$               & $1.142\pm0.075$               & $0.0337\pm0.0046$              & $ 0.0071\pm0.0128$              &    60.4\,/\,42 \\
    \hline
    \multirow{4}{*}{ ${}^{27}$Al }
    & \multicolumn{1}{c}{\bf 2.791} & \multicolumn{1}{c}{\bf 0.541} & \multicolumn{1}{c}{\bf 0.0130} & \multicolumn{1}{c|}{\bf 0.0363} &   212.6\,/\,45 \\
    & \multicolumn{1}{c}{\bf 2.791} & $0.628\pm0.008$               & \multicolumn{1}{c}{\bf 0.0130} & \multicolumn{1}{c|}{\bf 0.0363} &    80.9\,/\,43 \\
    & \multicolumn{1}{c}{\bf 2.791} & $0.660\pm0.010$               & $0.0102\pm0.0005$              & \multicolumn{1}{c|}{\bf 0.0363} &    49.7\,/\,42 \\
    & \multicolumn{1}{c}{\bf 2.791} & $0.653\pm0.028$               & $0.0098\pm0.0013$              & $ 0.0384\pm0.0081$              &    49.7\,/\,41 \\
    \hline
    \multirow{4}{*}{ ${}^{16}$O  }
    & \multicolumn{1}{c}{\bf 2.344} & \multicolumn{1}{c}{\bf 0.541} & \multicolumn{1}{c}{\bf 0.0120} & \multicolumn{1}{c|}{\bf 0.0348} &   123.9\,/\,48 \\
    & \multicolumn{1}{c}{\bf 2.344} & $0.558\pm0.026$               & \multicolumn{1}{c}{\bf 0.0120} & \multicolumn{1}{c|}{\bf 0.0348} &   123.6\,/\,47 \\
    & \multicolumn{1}{c}{\bf 2.344} & $0.637\pm0.022$               & $0.0057\pm0.0008$              & \multicolumn{1}{c|}{\bf 0.0348} &    58.7\,/\,46 \\
    & \multicolumn{1}{c}{\bf 2.344} & $0.601\pm0.067$               & $0.0036\pm0.0035$              & $ 0.0449\pm0.0180$              &    58.4\,/\,45 \\
    \hline
    \multirow{4}{*}{ ${}^{12}$C  }
    & \multicolumn{1}{c}{\bf 2.130} & \multicolumn{1}{c}{\bf 0.541} & \multicolumn{1}{c}{\bf 0.0120} & \multicolumn{1}{c|}{\bf 0.0348} &    85.5\,/\,13 \\
    & \multicolumn{1}{c}{\bf 2.130} & $0.666\pm0.010$               & \multicolumn{1}{c}{\bf 0.0120} & \multicolumn{1}{c|}{\bf 0.0348} &    10.4\,/\,11 \\
    & \multicolumn{1}{c}{\bf 2.130} & $0.571\pm0.046$               & $0.0094\pm0.0009$              & \multicolumn{1}{c|}{\bf 0.0348} &     6.4\,/\,10 \\
    & \multicolumn{1}{c}{\bf 2.130} & $0.521\pm0.083$               & $0.0070\pm0.0032$              & $ 0.0476\pm0.0185$              &     5.0\,/\,~9 \\
    \hline
    \multirow{4}{*}{ ${}^{2}$H  }
    & \multicolumn{1}{c}{\bf 1.172} & \multicolumn{1}{c}{\bf 0.541} & \multicolumn{1}{c}{\bf 0.0120} & \multicolumn{1}{c|}{\bf 0.0348} &    88.2\,/\,44 \\
    & \multicolumn{1}{c}{\bf 1.172} & $0.744\pm0.036$               & \multicolumn{1}{c}{\bf 0.0120} & \multicolumn{1}{c|}{\bf 0.0348} &    67.0\,/\,42 \\
    & \multicolumn{1}{c}{\bf 1.172} & $1.340\pm0.073$               & $0.0667\pm0.0080$              & \multicolumn{1}{c|}{\bf 0.0348} &    48.0\,/\,41 \\
    & \multicolumn{1}{c}{\bf 1.172} & $1.089\pm0.385$               & $0.0145\pm0.0051$              & $ 0.1452\pm0.0511$              &    45.8\,/\,40 \\
  \end{tabular}
\end{table*}
\endgroup

\section{Forward elastic $\boldsymbol{A_\text{N}^{pA}(t)}$ at $\boldsymbol{100\,\text{GeV}}$}

For the other nuclei shown in Fig.\,\ref{fig:pA}\,(left), the measured analyzing powers (see Table\,\ref{tab:Data}) are compared with the calculations in Fig.\,\ref{fig:Ru+}. Detailed results of the fits to the data are presented in Table\,\ref{tab:Ru}. In this study, we also consider the results for $p^\uparrow$C scattering (Table\,\ref{tab:DataC}), extracted from Fig.\,2 of Ref.\,\cite{Jinnouchi:2004up}.

For all nuclei, the charge density was fixed using the \emph{nominal values} of the Woods-Saxon parameters given in Table\,\ref{tab:WS}. Therefore, possible variations in $R_A$ and $a$ were considered in the fits only for the nuclear density distributions. However, because of the strong correlations between the free fit parameters, $R_A$ was always fixed at its \emph{nominal value} for light nuclei (Al, O, C, and $d$).

For all the nuclei considered, the measured analyzing powers are visually consistent with the parametrizations using the \emph{nominal values}, except for Ru and Zr at $-t>0.015\,\text{GeV}^2$. When the data are fitted using four free parameters for heavy nuclei or three for light nuclei, statistically acceptable agreement is achieved.

According to Table\,\ref{tab:Ru}, the \emph{nominal value} of $\operatorname{Im}r_5^{pA}$ is consistent with the results of fits in which all parameters are allowed to vary for all the nuclei considered, except Ru and Zr, for which discrepancies of approximately three standard deviations are observed.

However, for Zr, reducing the upper limit of the $t$-range from $0.019$ to $0.017\,\text{GeV}^2$---thereby excluding data points with limited statistical precision---yields $\chi^2/\text{ndf}$ values of 43.8/40 when $\operatorname{Im}r_5^{pA}$ is allowed to vary and 42.3/39 when it is fixed at the \emph{nominal value} of 0.0415. Thus, reducing the upper limit of $|t|$ substantially improves the fit, while fixing $\operatorname{Im}r_5^{p\text{Zr}}=0.0415$ has no significant effect on the $\chi^2$ value. A similar result is obtained for Ru by restricting the fit to $|t|<0.013\,\text{GeV}^2$.

It should also be noted that the analysis of the $^{96}_{44}$Ru data was performed using the charge density profile determined for $^{96}_{40}$Zr. However, if the charge density profile is adjusted based on the fit to the data, a regular fit over the range $|t|<0.019\,\text{GeV}^2$ yields $\chi^2/\text{ndf}$ values of 50.2/44 and 53.9/45 when $\operatorname{Im}r_5^{p\text{Ru}}$ is allowed to vary and fixed at its \emph{nominal value} of 0.0381, respectively. These results indicate that the \emph{nominal value} of the imaginary part of $r_5^{p\text{Ru}}$ is also consistent with the experimental data.

In Ref.\,\cite{Jinnouchi:2004up}, the proton beam polarization was known with a relative accuracy of approximately 10\%, and the $pC$ analyzing power was derived from the measured asymmetry using the central value of the polarization, $\langle P\rangle_\text{meas}$. With $R_A=2.13\,\text{fm}$ and $\operatorname{Im}r_5^{p\text{C}}=0.0348$ fixed in the fit, while $a$, $\operatorname{Re}r_5^{p\text{C}}$, and the correction factor to the beam polarization $C_P$ are allowed to vary, one obtains
\begin{equation}
  C_P=\langle P\rangle_\text{meas}/P_\text{beam}=0.985\pm0.021,
\end{equation}
where $P_\text{beam}$ is the beam polarization in the measurements reported in Ref.\,\cite{Jinnouchi:2004up}.

Thus, the model-dependent contribution to the uncertainty in the proton beam polarization determined using the $pC$ polarimeter can be estimated at 2\%. If the systematic corrections to $\operatorname{Re}r_5^{p\text{C}}$ can be reliably identified, the model-dependent systematic uncertainty of the $pC$ polarimeter may be reduced to 1\%.

For oxygen and carbon, the fits were also performed using modified harmonic-oscillator (MHO) parametrizations \cite{Wu:2025} for the nuclear density distributions. In neither case did this approach improve the fit. Similarly, using the Hulth\'en approximation \cite{Hulthen:1957} for the deuteron density distribution did not improve the fit. However, it should be emphasized that these results do not establish a preference for the Woods-Saxon parametrization for light nuclei until all necessary corrections to the model used to calculate the $pA$ analyzing power have been identified and accounted for.

\section{Summary}

A model for calculating the forward elastic analyzing power $A_\text{N}^{pA}(t)$, based on the Glauber approximation for $pA$ scattering and the calculation of soft-photon exchange corrections following Refs.\,\cite{Poblaguev:2025vtd, Poblaguev:2026wmb}, was tested against experimental data obtained at a beam energy of 100\,GeV in Refs.\,\cite{Poblaguev:2023stj,Jinnouchi:2004up}.

The input parameters were the Woods-Saxon parameters $R_A$ and $a$ of the nuclear density distribution and the proton-proton hadronic spin-flip amplitude parameter $r_5^{pp}$ \cite{Poblaguev:2019saw}. To determine the effective $r_5^{pA}$ \eqref{eq:r5pA} for proton-nucleus scattering, the proton-neutron value $r_5^{pn}$ at 100\,GeV is also needed because, at this energy, $r_5^{pn}\ne r_5^{pp}$ due to the large contribution of isovector Reggeons to $r_5$. The value of $r_5^{pn}$ was evaluated from a Regge fit to the HJET $pp$ data \cite{Poblaguev:2019saw}.

For the predefined, i.e., \emph{nominal}, values of the input parameters, the model-based calculations were found to be in reasonable, although only rough, agreement with the experimental data.

When the input parameters are treated as free parameters in the fit, the result is statistically consistent with the data.

Consequently, this approach may be used to obtain a sufficiently accurate parametrization of the measured $A_\text{N}^{pA}(t)$ in terms of four parameters for heavy nuclei and three for light nuclei. However, the resulting values of $R_A$, $a$, and $r_5^{pA}$ should not be interpreted as the actual values of the corresponding physical parameters. These can be determined reliably only after the unaccounted-for effects discussed in Section\,\ref{sec:pAu} have been accounted for.

Nevertheless, as already noted, the \emph{nominal values} of $\operatorname{Im}r_5^{pA}$ are in good agreement with the fit results for all the nuclei considered. With the caveat that the unaccounted-for effects may effectively alter the fitted values of $\operatorname{Im}r_5^{pA}$, we can conclude that the imaginary parts of the proton-proton hadronic spin-flip amplitudes measured at HJET \cite{Poblaguev:2019saw} are consistent with those observed in $p^\uparrow{A}$ scattering. If this interpretation is correct, the $p^\uparrow{A}$ analyzing power measurements considered here support the conclusion of Ref.\,\cite{Poblaguev:2019saw} regarding a non-zero Pomeron spin-flip coupling.

Some choices made in modeling the $pA$ analyzing power, such as using $\gamma_N^{pN}(b)$ rather than $\chi_N^{pN}(b)$ in Eq.\,\eqref{eq:gNpA} and using the Woods-Saxon parametrization for light nuclei, should be reconsidered after or during the elimination of the unaccounted-for effects.

As shown in Fig.\,\ref{fig:pA}\,(right), the $p^\uparrow$Au data acquired at HJET over a wide range of beam energies may allow a more detailed Regge fit of the proton-nucleon $r_5^{pN}(s)$ as a function of the squared center-of-mass energy than the fit performed in Ref.\,\cite{Poblaguev:2019saw}. The results of such a fit are important for pCarbon and $^3$He polarimetry at the EIC. However, the systematic uncertainties in the $p^\uparrow{A}$ measurements must first be reduced.

\section*{Acknowledgments}
The experimental part of this work, namely, the measurements of $pA$ analyzing power at HJET, was made possible largely through the efforts of Anatoli Zelenski and Grigor Atoian. The author would like to thank Boris Kopeliovich for numerous insightful discussions.
This work was authored by an employee of Brookhaven Science Associates, LLC, under Contract No. DE-SC0012704 with the U.S. Department of Energy.

\section*{DATA AVAILABILITY}

The data used in this work are provided in the text, figures, and tables of the paper.

\appendix

\section{Nuclear density distribution function \label{sec:WS}}

For heavy nuclei, the nuclear density distribution can be described by a two-parameter Fermi distribution \cite{Wu:2025}, also referred to as the Woods-Saxon parametrization:
\begin{equation}
\rho(r) = \frac{\rho_0}{1+\exp{\displaystyle\left(\frac{r-R_A}{a}\right)}}.
\end{equation}

The values of the parameters $R_A$ and $a$ for the nuclei considered in this paper are listed in Table\,\ref{tab:WS}. For $^{197}$Au, $^{96}$Zr, and $^{27}$Al, the Woods-Saxon parameters \cite{Wu:2025} were determined from low-energy electron-scattering experiments. For the other nuclei considered, the parameters were derived from these experimental values using simple relations, as explained in the table caption. Although the Woods-Saxon parametrization is not expected to be optimal for light nuclei such as $d$, C, and O, it was found to be adequate for a preliminary parametrization of the $p^\uparrow A$ analyzing powers studied here.

\begin{table}[t]
  \caption{ \label{tab:WS}
    Woods-Saxon nuclear density parameters for $^{197}$Au, $^{96}$Zr, and $^{27}$Al were taken from Ref.\,\cite{Wu:2025}. For $^{96}$Ru, the values of $R_A$ and $a$ for $^{96}$Zr were used. For the light nuclei $^{16}$O, $^{12}$C, and $^{2}$H ($d$), with atomic mass $A$, the parameters were derived from the $^{27}$Al values using $R_A = R_A^\text{(Al)}\times\left[A/A^\text{(Al)}\right]^{1/3}$ and $a=a^\text{(Al)}$.
  }
   \smallskip
  \begin{tabular}{ c | c c | c}
    \rule[-6pt]{0.pt}{0.5pt}
    & ~~$R_A\,\text{[fm]}~~$ & ~~$a\,\text{[fm]}~~$ & Reference \\
    \hline
    ${}^{197}$Au & 6.409 & 0.475 & \cite{Wu:2025} \\
    ${}^{ 96}$Ru & 4.995 & 0.507 & $^{96}$Zr       \\
    ${}^{ 96}$Zr & 4.995 & 0.507 & \cite{Wu:2025} \\
    ${}^{ 27}$Al & 2.791 & 0.541 & \cite{Wu:2025} \\
    ${}^{ 16}$O  & 2.344 & 0.541 & $^{27}$Al \\
    ${}^{ 12}$C  & 2.130 & 0.541 & $^{27}$Al \\
    ${}^{  2}$H $(d)$  & 1.172 & 0.541 & $^{27}$Al \\
  \end{tabular} 
\end{table}

\begingroup
\setlength{\tabcolsep}{12pt}
\begin{table}[t]
  \caption{ \label{tab:r5A}
    Estimates of the proton-nucleus spin-flip amplitude parameters $r_5^{pA}$ for the considered neclei. 
  }
  \smallskip
  \begin{tabular}{ r | r r }
  \rule[-6pt]{0.pt}{0.5pt}  & \multicolumn{1}{|c}{$\operatorname{Re}r_5$}
    & \multicolumn{1}{ c}{$\operatorname{Im}r_5$} \\
     \hline
    $p^\uparrow p$  & $-$0.0164 & $-$0.0053 \\
    $p^\uparrow n$  &  0.0404 &  0.0749 \\
    \hline
    $p^\uparrow d$,  $p^\uparrow$C, $p^\uparrow$O   &  0.0120 &  0.0348 \\
    $p^\uparrow$Al  &  0.0130 &  0.0363 \\
    $p^\uparrow$Zr  &  0.0167 &  0.0415 \\
    $p^\uparrow$Ru  &  0.0144 &  0.0381 \\
    $p^\uparrow$Au  &  0.0176 &  0.0427  
  \end{tabular}
\end{table}
\endgroup

\begingroup
\setlength{\tabcolsep}{5pt}
\begin{table*}[t]
  \caption{ \label{tab:Data}
    Preliminary results for the analyzing power measurements at HJET for 100\,GeV beam energy. The momentum transfer squared $t$ was derived from the experimentally determined recoiled proton energy $T_R$.
  }
  \smallskip
  \small
  \begin{tabular}{ c | c | r r r r r r }
    $\boldsymbol{T_R}$ & $\boldsymbol{-t=2m_pT_R}$ & \multicolumn{6}{|c}{ $\boldsymbol{A_\text{N}^{pA}(t)\times10^3}$} \\
    \text{[MeV]} & [GeV${}^2$] 
    & \multicolumn{1}{c}{${}^{197}$Au  }
    & \multicolumn{1}{c}{${}^{96}$Ru  }
    & \multicolumn{1}{c}{${}^{96}$Zr  }
    & \multicolumn{1}{c}{${}^{27}$Al  }
    & \multicolumn{1}{c}{${}^{16}$O  }
    & \multicolumn{1}{c}{${}^{2}$H  }
    \\
  \hline
  0.70 &  0.00131  & $ 32.99\pm 0.11$  & $ 34.60\pm 0.28$  & $ 34.07\pm 0.27$  & $ 34.03\pm 0.24$  & $ 32.25\pm 0.47$  & $ 24.45\pm 3.12$  \\  
  0.90 &  0.00169  & $ 33.92\pm 0.13$  & $ 37.47\pm 0.33$  & $ 36.05\pm 0.31$  & $ 36.91\pm 0.25$  & $ 36.54\pm 0.53$  & $ 27.98\pm 2.80$  \\ 
  1.10 &  0.00206  & $ 31.52\pm 0.14$  & $ 37.66\pm 0.35$  & $ 35.48\pm 0.33$  & $ 37.28\pm 0.26$  & $ 35.57\pm 0.52$  & $ 24.11\pm 2.73$  \\ 
  1.30 &  0.00244  & $ 27.71\pm 0.17$  & $ 34.48\pm 0.40$  & $ 32.84\pm 0.37$  & $ 35.65\pm 0.29$  & $ 35.24\pm 0.58$  & $ 24.36\pm 2.37$  \\ 
  1.50 &  0.00281  & $ 23.30\pm 0.18$  & $ 32.40\pm 0.41$  & $ 30.97\pm 0.38$  & $ 35.00\pm 0.30$  & $ 33.30\pm 0.58$  & $ 23.69\pm 2.55$  \\ 
  1.70 &  0.00319  & $ 18.63\pm 0.20$  & $ 29.20\pm 0.42$  & $ 27.83\pm 0.39$  & $ 33.22\pm 0.30$  & $ 33.25\pm 0.58$  & $ 22.17\pm 2.21$  \\ 
  1.90 &  0.00357  & $ 13.72\pm 0.23$  & $ 27.13\pm 0.43$  & $ 24.82\pm 0.40$  & $ 32.08\pm 0.30$  & $ 31.06\pm 0.58$  & $ 20.19\pm 2.46$  \\ 
  2.10 &  0.00394  & $  9.30\pm 0.29$  & $ 24.86\pm 0.52$  & $ 21.49\pm 0.48$  & $ 29.82\pm 0.35$  & $ 29.72\pm 0.67$  & $ 23.49\pm 2.30$  \\ 
  2.30 &  0.00432  & $  5.57\pm 0.31$  & $ 21.12\pm 0.54$  & $ 18.89\pm 0.50$  & $ 29.78\pm 0.35$  & $ 27.85\pm 0.67$  & $ 18.87\pm 1.91$  \\ 
  2.50 &  0.00469  & $  0.57\pm 0.28$  & $ 19.32\pm 0.48$  & $ 16.39\pm 0.45$  & $ 27.36\pm 0.31$  & $ 26.65\pm 0.59$  & $ 19.42\pm 2.17$  \\ 
  2.70 &  0.00507  & $ -4.56\pm 0.35$  & $ 15.63\pm 0.58$  & $ 14.01\pm 0.54$  & $ 26.17\pm 0.36$  & $ 26.14\pm 0.69$  & $ 22.56\pm 2.20$  \\ 
  2.90 &  0.00544  & $ -9.09\pm 0.36$  & $ 14.40\pm 0.61$  & $ 11.70\pm 0.57$  & $ 24.60\pm 0.37$  & $ 26.31\pm 0.69$  & $ 18.42\pm 2.64$  \\ 
  3.10 &  0.00582  & $-12.76\pm 0.46$  & $ 10.49\pm 0.77$  & $  9.91\pm 0.72$  & $ 23.74\pm 0.45$  & $ 25.31\pm 0.84$  & $ 15.32\pm 2.18$  \\ 
  3.30 &  0.00619  & $-17.23\pm 0.40$  & $  8.86\pm 0.65$  & $  7.62\pm 0.61$  & $ 22.84\pm 0.37$  & $ 23.24\pm 0.69$  & $ 20.71\pm 2.17$  \\ 
  3.50 &  0.00657  & $-21.76\pm 0.44$  & $  6.13\pm 0.68$  & $  5.50\pm 0.64$  & $ 21.59\pm 0.38$  & $ 23.16\pm 0.70$  & $ 16.41\pm 2.70$  \\ 
  3.70 &  0.00694  & $-25.64\pm 0.58$  & $  6.04\pm 0.88$  & $  3.78\pm 0.82$  & $ 20.76\pm 0.46$  & $ 22.45\pm 0.86$  & $ 17.19\pm 2.14$  \\ 
  3.90 &  0.00732  & $-30.21\pm 0.51$  & $  1.54\pm 0.74$  & $  1.75\pm 0.70$  & $ 20.08\pm 0.38$  & $ 21.49\pm 0.71$  & $ 15.81\pm 2.61$  \\ 
  4.10 &  0.00769  & $-33.48\pm 0.68$  & $  0.05\pm 0.95$  & $ -0.61\pm 0.90$  & $ 18.81\pm 0.48$  & $ 21.28\pm 0.87$  & $ 15.69\pm 2.10$  \\ 
  4.30 &  0.00807  & $-36.18\pm 0.61$  & $ -2.94\pm 0.82$  & $ -3.12\pm 0.77$  & $ 17.79\pm 0.39$  & $ 19.34\pm 0.72$  & $ 14.50\pm 2.53$  \\ 
  4.50 &  0.00844  & $-38.84\pm 0.82$  & $ -6.05\pm 1.06$  & $ -4.48\pm 0.99$  & $ 17.08\pm 0.49$  & $ 19.14\pm 0.89$  & $ 18.12\pm 2.04$  \\ 
  4.70 &  0.00882  & $-41.72\pm 0.74$  & $ -6.84\pm 0.91$  & $ -7.31\pm 0.86$  & $ 15.66\pm 0.40$  & $ 18.07\pm 0.73$  & $ 20.53\pm 2.49$  \\ 
  4.90 &  0.00920  & $-44.36\pm 1.02$  & $ -8.75\pm 1.17$  & $ -8.56\pm 1.11$  & $ 14.94\pm 0.50$  & $ 19.18\pm 0.91$  & $ 11.85\pm 2.47$  \\ 
  5.10 &  0.00957  & $-42.74\pm 1.12$  & $-13.60\pm 1.24$  & $-11.84\pm 1.17$  & $ 14.11\pm 0.51$  & $ 17.24\pm 0.92$  & $ 19.92\pm 2.50$  \\ 
  5.30 &  0.00995  & $-42.64\pm 1.22$  & $-11.21\pm 1.31$  & $-13.80\pm 1.24$  & $ 14.70\pm 0.52$  & $ 16.28\pm 0.92$  & $ 17.33\pm 2.46$  \\ 
  5.50 &  0.01032  & $-33.74\pm 1.33$  & $-16.04\pm 1.35$  & $-16.16\pm 1.28$  & $ 13.73\pm 0.52$  & $ 18.13\pm 0.93$  & $ 10.68\pm 2.48$  \\ 
  5.70 &  0.01070  & $-29.05\pm 1.44$  & $-19.24\pm 1.43$  & $-17.99\pm 1.36$  & $ 12.30\pm 0.53$  & $ 15.79\pm 0.93$  & $ 16.40\pm 2.46$  \\ 
  5.90 &  0.01107  & $-18.11\pm 1.55$  & $-21.10\pm 1.50$  & $-22.50\pm 1.43$  & $ 12.11\pm 0.54$  & $ 15.62\pm 0.94$  & $ 16.04\pm 2.52$  \\ 
  6.10 &  0.01145  & $ -9.11\pm 1.67$  & $-22.08\pm 1.60$  & $-20.89\pm 1.53$  & $ 11.17\pm 0.55$  & $ 15.88\pm 0.96$  & $ 13.69\pm 2.52$  \\ 
  6.30 &  0.01182  & $  4.32\pm 1.73$  & $-23.17\pm 1.71$  & $-24.01\pm 1.64$  & $ 11.93\pm 0.57$  & $ 15.08\pm 0.98$  & $ 17.86\pm 2.58$  \\ 
  6.50 &  0.01220  & $ 20.09\pm 1.83$  & $-24.93\pm 1.81$  & $-26.18\pm 1.75$  & $ 10.95\pm 0.60$  & $ 15.35\pm 1.00$  & $ 20.33\pm 2.61$  \\ 
  6.70 &  0.01257  & $ 28.12\pm 1.88$  & $-29.52\pm 1.93$  & $-25.77\pm 1.88$  & $  9.38\pm 0.61$  & $ 15.36\pm 1.06$  & $ 10.34\pm 2.68$  \\ 
  6.90 &  0.01295  & $ 38.21\pm 1.92$  & $-30.45\pm 2.06$  & $-28.24\pm 2.00$  & $  9.44\pm 0.66$  & $ 15.33\pm 1.12$  & $ 15.07\pm 2.73$  \\ 
  7.10 &  0.01332  & $ 43.25\pm 1.93$  & $-31.13\pm 2.22$  & $-34.62\pm 2.19$  & $ 10.40\pm 0.69$  & $ 13.09\pm 1.18$  & $ 14.16\pm 2.62$  \\ 
  7.30 &  0.01370  & $ 48.22\pm 1.86$  & $-31.71\pm 2.41$  & $-33.05\pm 2.29$  & $  8.78\pm 0.70$  & $ 11.90\pm 1.23$  & $ 16.52\pm 2.55$  \\ 
  7.50 &  0.01407  & $ 51.90\pm 1.73$  & $-37.24\pm 2.66$  & $-34.01\pm 2.44$  & $  7.55\pm 0.67$  & $ 11.02\pm 1.22$  & $ 10.06\pm 2.50$  \\ 
  7.70 &  0.01445  & $ 53.19\pm 1.60$  & $-38.22\pm 3.03$  & $-37.82\pm 2.75$  & $  7.64\pm 0.64$  & $  9.11\pm 1.40$  & $ 12.47\pm 3.63$  \\ 
  7.90 &  0.01482  & $ 53.93\pm 1.81$  & $-40.70\pm 3.30$  & $-38.15\pm 3.02$  & $  7.10\pm 0.70$  & $  9.99\pm 1.02$  & $ 14.01\pm 3.39$  \\ 
  8.10 &  0.01520  & $ 49.75\pm 2.20$  & $-36.57\pm 4.19$  & $-33.78\pm 3.84$  & $  5.97\pm 0.89$  & $ 12.01\pm 1.60$  & $  9.20\pm 3.53$  \\ 
  8.30 &  0.01558  & $ 51.66\pm 2.24$  & $-36.22\pm 4.64$  & $-39.97\pm 4.26$  & $  5.46\pm 0.91$  & $  7.75\pm 1.47$  & $  7.59\pm 3.94$  \\ 
  8.50 &  0.01595  & $ 50.85\pm 2.38$  & $-39.29\pm 5.38$  & $-36.86\pm 4.90$  & $  4.32\pm 1.02$  & $  7.68\pm 1.50$  & $ 17.13\pm 6.26$  \\ 
  8.70 &  0.01633  & $ 49.19\pm 3.60$  & $-41.20\pm 8.44$  & $-37.41\pm 7.65$  & $  4.95\pm 1.60$  & $ 13.11\pm 2.22$  & $ 22.57\pm 5.04$  \\ 
  8.90 &  0.01670  & $ 47.36\pm 2.76$  & $-46.81\pm 6.64$  & $-40.36\pm 6.23$  & $  5.90\pm 1.22$  & $  6.83\pm 1.66$  & $ 17.89\pm 6.39$  \\ 
  9.10 &  0.01708  & $ 42.12\pm 3.15$  & $-42.07\pm 8.34$  & $-22.24\pm 7.56$  & $  3.38\pm 1.35$  & $  8.09\pm 1.79$  & $ 17.71\pm11.64$  \\ 
  9.30 &  0.01745  & $ 44.26\pm 5.46$  & $-47.75\pm14.93$  & $ -5.68\pm13.70$  & $  4.62\pm 2.23$  & $  9.11\pm 2.85$  & $ 20.59\pm10.28$  \\ 
  9.50 &  0.01783  & $ 42.82\pm 4.81$  & $-52.72\pm13.57$  & $-18.14\pm12.39$  & $  5.10\pm 1.88$  & $  6.62\pm 2.26$  & $ -8.16\pm19.16$  \\ 
  9.70 &  0.01820  & $ 33.19\pm 8.59$  & $-10.99\pm19.37$  & $-19.68\pm17.19$  & $ -3.91\pm 2.94$  & $  4.93\pm 2.69$  & $ 35.84\pm47.91$  \\ 
  9.90 &  0.01858  &                   & $ 35.86\pm34.48$  & $-15.79\pm29.74$  &                   & $  4.19\pm 4.19$  & $-71.01\pm51.15$  \\ 
% 10.10 &  0.01895  &                   & $ 19.70\pm26.89$  & $ 17.88\pm25.26$  &                   & $  4.28\pm 3.32$  & $-174.47\pm183.43$\\
% 10.30 &  0.01933  &                   & $-97.83\pm62.04$  & $-70.86\pm48.75$  &                   & $ 21.34\pm 5.56$  &                   \\
% 10.50 &  0.01970  &                   & $ 18.91\pm58.11$  & $ 40.51\pm52.21$  &                   & $ 11.10\pm 5.14$  &                   \\
  \end{tabular}
\end{table*}
\endgroup

\begingroup
\setlength{\tabcolsep}{8pt}
\begin{table*}[t]
  \caption{ \label{tab:DataC}
    Results for the measured $A_\text{N}(t)$ at 100\,GeV as extracted from Fig.\,2 of Ref.\,\cite{Jinnouchi:2004up}. In the analysis presented here, the point at $-t=0.00478\,\text{GeV}^2$ was not considered.
  }
 % \smallskip
  \small
  \begin{tabular}{ c  r | c r | c r }
%    \rule[12pt]{0.pt}{0.5pt} \\
    \rule[-6pt]{0.pt}{0.5pt}%
    $-t$\;\;[GeV${}^2$] & \multicolumn{1}{c|}{$A_\text{N}(t)$} &
    $-t$\;\;[GeV${}^2$] & \multicolumn{1}{c|}{$A_\text{N}(t)$} &
    $-t$\;\;[GeV${}^2$] & \multicolumn{1}{c }{$A_\text{N}(t)$} \\
    \hline
 0.00781 &  $ 0.01927\pm 0.00027$ &  0.01895  & $ 0.00757\pm 0.00035$ &  0.03116  & $ 0.00003\pm 0.00050$ \\
 0.01005 &  $ 0.01605\pm 0.00027$ &  0.02122  & $ 0.00538\pm 0.00035$ &  0.03562  & $-0.00254\pm 0.00062$ \\
 0.01225 &  $ 0.01300\pm 0.00027$ &  0.02342  & $ 0.00367\pm 0.00041$ &  0.04105  & $-0.00565\pm 0.00059$ \\
 0.01451 &  $ 0.01073\pm 0.00027$ &  0.02565  & $ 0.00284\pm 0.00047$ &  0.04780  & $-0.00606\pm 0.00074$ \\
 0.01675 &  $ 0.00887\pm 0.00027$ &  0.02790  & $ 0.00219\pm 0.00053$ &           &                       \\ 
  \end{tabular}
\end{table*}
\endgroup

\section{Evaluation of $\boldsymbol{r_5^{pA}}$   \label{sec:r5} }

In Ref.\,\cite{Poblaguev:2019saw}, the dependence of the elastic proton-proton spin-flip parameter $r_5$ on the squared center-of-mass energy $s$ was evaluated using a Regge fit to the measured values of $r_5$:
\begin{equation}
  \sigma_\textrm{tot}(s)\times r_5^{pp}(s) = \sum_{\mathcal{R}=P,R^\pm}{f_5^\mathcal{R}\,\mathcal{R}(s)}, 
  \label{eq:r5Regge}
\end{equation}
where the Reggeon and Pomeron components $\mathcal{R}(s)$ were defined in Ref.\,\cite{Fagundes:2017iwb} using the AU-L$\gamma$=2(T) fit to the non-flip $pp$ amplitudes:
\begin{equation}
  \sigma_\textrm{tot}(s)\times\left[\rho(s)+i\right] = \sum_{\mathcal{R}=P,R^\pm}\mathcal{R}(s).
  \label{eq:Regge}
\end{equation}

Treating the couplings $f_5^\mathcal{R}$ as free parameters in a combined fit to the HJET data at $100$ and $255$\,GeV \cite{Poblaguev:2019saw}, the following values were obtained:
\begin{eqnarray}
  f_5^{P^{\phantom+}} & = & {\phantom-}0.045 \pm 0.002_\textrm{stat} \pm 0.003_\textrm{syst},
  \label{eq:f5P}\\
  f_5^{R^+}                 & = & -0.032\pm 0.007_\textrm{stat} \pm 0.014_\textrm{syst}, \label{eq:f5R+}\\
  f_5^{R^-}                 & = & {\phantom-}0.622\pm 0.023_\textrm{stat} \pm 0.024_\textrm{syst}.
  \label{eq:f5R-} 
\end{eqnarray}

Using these values, the proton-proton $r_5$ at 100\,GeV can be decomposed into contributions from the isoscalar $(R^+)$ and isovector $(R^-)$ Reggeons and the Pomeron $P$:
\begin{align}
  r_5^{R^-}        &\approx          -0.031-i\,0.037, \\
  r_5^{R^+}        &\approx\phantom{-}0.005-i\,0.008, \\
  r_5^{P^{\phantom{+}}} &\approx\phantom{-}0.005+i\,0.046.
\end{align}

To calculate the proton-neutron hadronic spin-flip amplitude, the sign of the isovector contribution must be reversed, yielding
\begin{equation}
  r_5^{pn} = r_5^{P} +  r_5^{R^+} - r_5^{R^-} = 0.0404 + i\,0.0749.
\end{equation}
It should be noted that, because the isovector spin-flip coupling $f_5^{R^-}$ is very large compared with $f_5^{R^+}$ and $f_5^{P}$, the substantial difference between $r_5^{pp}$ and $r_5^{pn}$ is consistent with the approximation $f_N^{pn}(q)\approx f_N^{pp}(q)$ for the non-flip amplitudes, which is conventionally assumed in the Glauber approach.

Estimates for all nuclei considered in this work, based on Eq.\,\eqref{eq:r5pA}, are summarized in Table\,\ref{tab:r5A}.

\section{The Coulomb correction function $\boldsymbol{\widetilde{\mathcal{F}}(q^2)}$   \label{sec:CC} }

In this section, the transverse momentum $q$ and impact parameter $b$ are rescaled to dimensionless variables:
\begin{equation}
  \sqrt{\frac{B_C}{2}}q\to q;\quad \sqrt{\frac{2}{B_C}}b\to b.
  \label{eq:dimless}
\end{equation}
In terms of these dimensionless variables, the Coulomb correction $\widetilde{\mathcal{F}}(q^2)$ can be calculated \cite{Poblaguev:2025vtd} as
\begin{flalign}
  &\widetilde{\mathcal{F}}(q^2) =%
  e^{q^2} \times  \Bigg[
  e^{-i\alpha{Z}\ln{q^2}}\,%
  \frac{\Gamma\left(1 + i\alpha{Z}\right)}{\Gamma\left(1 - i\alpha{Z}\right)}
  \nonumber \\&
  + q \int_0^{20} \!\! db \, e^{i\alpha{Z}\left[\ln{(b^2/4)} + E_1(b^2/4)\right]}%
  \left(1\!-\!e^{-b^2/4}\right) J_1(qb)
  \nonumber \\ &-
  \begin{cases}
    \Sigma_R(20q)        & \text{if}~ q\leq0.1,\\[7pt]
    \Sigma_R(2)+ q\int_2^{20}{ \!db\,e^{i\alpha{Z}\ln{(b^2/4)}}J_1(qb)} & \text{if}~ q>0.1,
  \end{cases}
  \Bigg]
  \label{eq:FM}
\end{flalign}
where
\begin{equation}
  \Sigma_R(a) = \frac{a^2}{4}%
  \sum_{k=0}^\infty{\frac{(-a^2/4)^k}{k!(k+1)!}\,%
    \frac{ e^{i\alpha{Z}\ln{(a^2/4q^2)}}}{k+1+i\alpha{Z}}
  }.
  \label{eq:Sigma}
\end{equation}

The series in Eq.\,\eqref{eq:Sigma} converges rapidly for $a\!<\!2$, allowing it to be evaluated with high accuracy. The integrals appearing explicitly in Eq.\,\eqref{eq:FM} can be evaluated numerically for any reasonable value of $Z$ using uniformly spaced sampling points; thus, their numerical evaluation poses no significant difficulties.

Notably, for fixed $Z$, $\widetilde{\mathcal{F}}(q^2)$ depends only on the dimensionless parameter $q$ defined in Eq.\,\eqref{eq:dimless}. Therefore, $\widetilde{\mathcal{F}}(q^2)$ can be tabulated in advance to the required precision before the Coulomb corrections are calculated.

\section{Experimental $\boldsymbol{A_\text{N}^{pA}(t)}$ data used \label{sec:data} }

In this work, the model-based calculations are tested against \emph{unpublished} experimental data obtained with a 100\,GeV beam (in the laboratory frame) at RHIC HJET \cite{Poblaguev:2023stj}, as well as measurements performed using an ultra-thin carbon target \cite{Jinnouchi:2004up}.

The $A_\text{N}^{pA}(t)$ data for six nuclei measured at HJET are presented in Table\,\ref{tab:Data}.

The $p$C data extracted from Fig.\,2 of Ref.\,\cite{Jinnouchi:2004up} are presented in Table\,\ref{tab:DataC}. The measurements cover a wider range of momentum transfer, $0.007<-t<0.048\,\text{GeV}^2$. It should be noted that the data point at $t=-0.0478\,\text{GeV}^2$ was not included in the present analysis.

\bibliographystyle{apsrev4-2}
%\bibliography{_pA.bib}

\end{document}